\documentclass[%
onecolumn,
12pt]{revtex4-2}

\usepackage{graphicx}% Include figure files
\usepackage{dcolumn}% Align table columns on decimal point
\usepackage{bm}% bold math
\usepackage{amsmath}
\usepackage{cases}
\usepackage{mathrsfs}
\usepackage{array}  
\usepackage{booktabs}
\usepackage{float}
\usepackage{multirow}
\usepackage{enumerate}
\usepackage{hyperref}% add hypertext capabilities
\hypersetup{colorlinks=true,linkcolor=blue,anchorcolor=blue,citecolor=blue}
\begin{document}

\title{Instability and Turbulent Relaxation in a Stochastic Magnetic Field}% Force line breaks with \\
\author{Mingyun Cao}
\author{P. H. Diamond}%
\email{pdiamond@ucsd.edu}
\affiliation{%
 Department of Physics, University of California, San Diego
}%

\date{November 8, 2021}% It is always \today, today,
             %  but any date may be explicitly specified

\begin{abstract}
An analysis of instability dynamics in a stochastic magnetic field is presented for the tractable case of the resistive interchange. Externally prescribed static magnetic perturbations convert the eigenmode problem to a stochastic differential equation, which is solved by the method of averaging. The dynamics are rendered multi-scale, due to the size disparity between the test mode and magnetic perturbations. Maintaining quasi-neutrality at all orders requires that small-scale convective cell turbulence be driven by disparate scale interaction. The cells in turn produce turbulent mixing of vorticity and pressure, which is calculated by fluctuation-dissipation type analyses, and are relevant to pump-out phenomena. The development of correlation between the ambient magnetic perturbations and the cells is demonstrated, showing that turbulence will `lock on' to ambient stochasticity. Magnetic perturbations are shown to produce a magnetic braking effect on vorticity generation at large scale. Detailed testable predictions are presented. The relations of these findings to the results of available simulations and recent experiments are discussed.
\end{abstract}

%\keywords{Suggested keywords}%Use showkeys class option if keyword
                              %display desired
\maketitle

%\tableofcontents
\section{\label{introduction}Introduction}
The dynamics of instability, relaxation, and turbulence are (taken collectively) fundamental to magnetic confinement physics. Here, `relaxation' includes the evolution of plasma free energy (in the presence of sources and sinks), and the resulting transport~\cite{PhysRevLett.33.1139}. Relaxation determines plasma confinement and possible bifurcations between different states thereof~\cite{wagner2007quarter}. Recently, a new element has been added to this already challenging problem. Good confinement is no longer deemed sufficient. Rather, good confinement must be achieved along with good power handling and boundary control~\cite{Kikuchi2019}. Hence, plasma relaxation must be addressed in a base state which is either three dimensional or even stochastic. A specific example of this is the Resonant Magnetic Perturbation, or RMP~\cite{evans2006edge}. The development of RMP was motivated by the desire to mitigate or suppress ELM-driven relaxation by inducing a stochastic layer at the plasma edge. The hope was that mitigation could be achieved without excessive degradation of confinement. One consequence of inducing such extrinsic stochasticity is that turbulence evolution and transport bifurcation now occur in a background with chaotic magnetic fields, and so the theory must address this. In the case of RMP plasmas, models of pedestal transport~\cite{Eric2012}, the L-H transition~\cite{Samantha2021}, flow and electric field shear evolution~\cite{Weixin2021}, and turbulence dynamics~\cite{Samantha2020} all must be re-formulated to account for the presence of extrinsic stochasticity and its effects. These problems present many challenges, starting with the need to revisit fundamental instability dynamics in a stochastic background. That problem constitutes the primary motivation for this paper. We note here, in passing, that the case of RMP is not unique. Stellarator confinement~\cite{W7X2005}, magnetic island evolution~\cite{Ida2016}, and the dynamics of disruptions~\cite{Boozer2012} all require us to confront the coexistence and synergy of instability, turbulence and magnetic stochasticity.

As noted above, the very first question we must answer is how ambient magnetic stochasticity impacts instability evolution. This is something of a classic problem in MFE theory, and early interest in it was motivated by the persistence of MHD-like phenomena in high temperature plasmas, where decoupling of field and fluid by resistivity was ineffective. Magnetic braiding by stochastic fields---which was hypothesized to produce electron heat transport~\cite{rechrosen}, or, equivalently, electron viscosity~\cite{Itoh1994}---was a natural alternative candidate. Of this genre of work, the paper of Kaw, et al.~\cite{Kaw1979} is especially well known. This analysis invoked anomalous electron viscosity to trigger tearing mode growth. That calculation followed the idea of `low-$\boldsymbol{k}$ mode meets high-$\boldsymbol{k}$ ambient background'---i.e., a problem in disparate scale interaction of a low-$\boldsymbol{k}$ coherent fluctuation (single mode) with high-$\boldsymbol{k}$ turbulence. Several other works pursued and developed the electron viscosity/hyper-resistivity idea~\cite{Strauss1986,Bhatta&Hameiri1986,Xu2010}. All such papers focused on magnetic stochasticity-as-anomalous-dissipation, and did not address relevant issues such as stochasticity effects on mode structure (i.e. the large-scale mode lives in an effective potential which is random), self-consistency effects, and closure of the microscale $\leftrightarrow$ macroscale feedback loop. All treated the effects of stochasticity using a quasilinear-type approach in which quasi-neutrality was not maintained at all orders in the analysis. More generally, these analyses were not systematic.

A clue to the importance of maintaining $\nabla\cdot\boldsymbol{J}=0$ at all orders of the calculation may be found in the theory of stochastic field induced heat transport proposed by Kadomtsev and Pogutse~\cite{K&P1979}. There, $\nabla\cdot\boldsymbol{q}=0$ was maintained throughout the analysis, and forced consideration of static temperature fluctuations on small scales, which were induced by the imposed magnetic perturbations. Temperature fluctuations are generated by the interaction of magnetic perturbations and the mean temperature profile. The effects of these temperature fluctuations was to cause a dramatic reduction in the effective cross-field heat conductivity, due to a cancellation between leading terms in the heat flux. The message was clear ---maintaining $\nabla\cdot\boldsymbol{q}=0$ revealed the importance of considering accompanying small-scale temperature perturbations induced by $\boldsymbol{\tilde{b}}$, which in turn forced a significant departure from kinematic expectations. In the problem considered here, $\boldsymbol{\tilde{b}}$ induces small-scale \emph{potential} fluctuations, which have important effects.

\begin{figure}[h]
\includegraphics[scale=0.65]{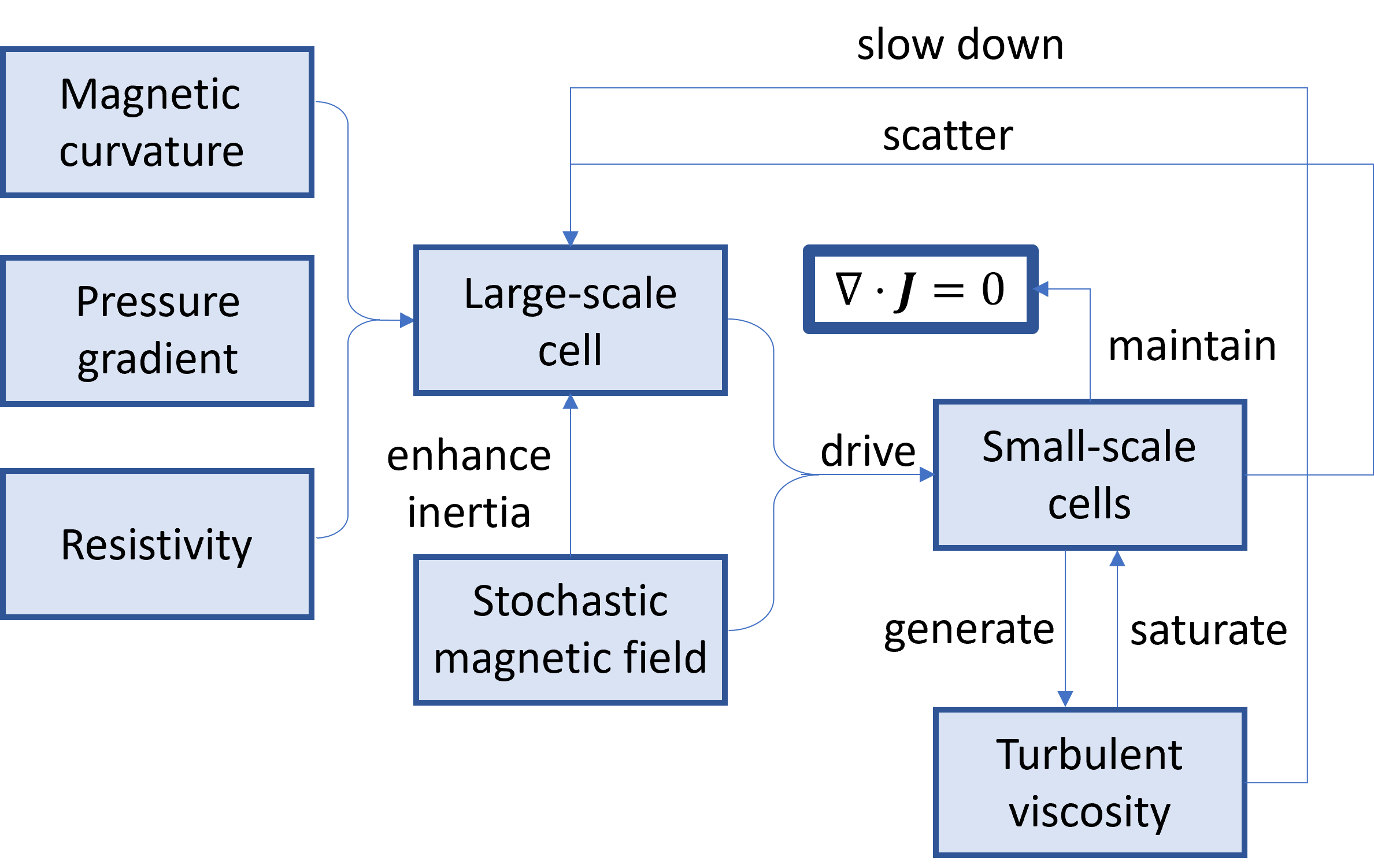}
\caption{\label{flowchart} Multi-scale feedback loops: At large scale, there is a single resistive interchange mode driven by magnetic curvature and mean pressure gradient. When a stochastic magnetic field is prescribed, to maintain $\nabla\cdot\boldsymbol{J}=0$ at all scales, the beat of the large-scale cell and the stochastic magnetic field can drive small-scale convective cells, which further generate a turbulent viscosity and a turbulent diffusivity. The effects of stochastic magnetic field on the large-scale cell are: 1, stochastic magnetic field itself can enhance the plasma inertia and then oppose the mode growth---i.e., magnetic braking effect; 2, the resultant turbulent viscosity can slow down the mode growth; 3, small-scale convective cells can modify large-scale cell via electrostatic scattering. The growth of small-scale convective cells is saturated by the turbulent viscosity, and adiabatically modulated by the beat mentioned above. Thus large scale and small scale are interacted.}

\end{figure}

In this paper, we present the theory of a simple instability in a static, ambient stochastic magnetic field. The instability studied is the electrostatic resistive interchange, and this choice is motivated by the desire for simplicity. The problem is framed as one where we seek to determine the evolution of a particular low-$m$ mode in a fixed, stochastic background. However, maintenance of quasi-neutrality throughout the analysis brings a surprise! We show that the interaction of the imposed magnetic perturbations and the large-scale structure must necessarily drive a spectrum of small-scale convective cells. \emph{Thus, what one thinks of as a problem of a single electrostatic mode in a stochastic background is actually a multi-scale convective cell turbulence problem!} We see that small-scale magnetic stochasticity drives small-scale convective turbulence as a consequence of quasi-neutrality. Enhanced transport---represented by a turbulent viscosity and thermal diffusivity---results. The analysis employs the method of averaging to derive coupled small-scale fluctuations and mean field (i.e., large-scale cell) equations. Thus the analysis has features in common with that for multi-scale problems~\cite{Diamond2005,Rame&PatPPCF,McDevitt2013,Ishizawa2007}. An interesting finding of this calculation is that small scales exert a magnetic braking effect on large scales. This effect resembles---but is not identical---to the magnetic braking predicted for tearing modes by Rutherford~\cite{Rutherford1973}. The analysis incorporates multi-scale feedback loops, which couple the dynamics of the large-scale envelope and small-scale cells. The structure of the analysis is shown in FIG.~\ref{flowchart}. We discuss the relation of our results to previous simulations and current experiments. 

The remainder of this paper is organized as follows. We first elaborate on the construction of our model in Sec.~\ref{model}. Quantitative results, including the stochasticity induced correction to the growth of the large-scale mode, the scaling of the turbulent viscosity $\nu$, and the correlation $\langle\tilde{b}_{r}\tilde{v}_{r}\rangle$, are also given in Sec.~\ref{model}. In Sec.~\ref{analysis}, we discuss the physical interpretations of the growth rate correction, with an emphasis on the magnetic braking effect. Conclusion and discussion are in Sec.~\ref{conclusion}. Exact solutions of the eigenmode equation Eq.(\ref{vorticity}), and a brief introduction for Kadomtsev and Pogutse's 1979 work are provided in the Appendix.
\section{\label{model}Model Development}
Using the quasi-neutrality condition, a model for a large-scale single resistive interchange mode in a stochastic field background is derived and developed in this section. We start with a model of resistive interchange instability driven by magnetic curvature and mean pressure gradient in a smooth magnetic field, and then introduce the stochastic magnetic field by rewriting the parallel gradient operator with the random bending effect included---i.e., $\nabla_{\|}\rightarrow\nabla_{\|}^{(0)}+\boldsymbol{\tilde{b}}\cdot\nabla_{\bot}$. But as will be discussed later, the effects of a stochastic magnetic field are more involved than that. To maintain quasi-neutrality, i.e., $\nabla\cdot\boldsymbol{J}=0$, small-scale convective cells must be generated. These constitute intrinsically multi-scale microturbulence. The basic equations in our model are vorticity equation, pressure equation, and electrostatic Ohm's law of resistive MHD
\begin{equation}
\bigg(\frac{\partial}{\partial t}+\tilde{\boldsymbol{v}}\cdot\nabla\bigg) \nabla_{\perp}^{2}(\bar{\varphi}+\tilde{\varphi})=
\frac{\eta S}{\tau_{A}}\nabla_{\|}J_{\|}-\frac{\kappa B_{0}}{\rho_{0}} \frac{\partial\left(\bar{p}_{1}+\tilde{p}_{1}\right)}{\partial y},
\label{modifiedvorticity}
\end{equation}
\begin{equation}
\left(\frac{\partial}{\partial t}+\tilde{\boldsymbol{v}}\cdot\nabla\right)\left(\bar{p}_{1}+\tilde{p}_{1}\right)-\frac{\nabla(\bar{\varphi}+\tilde{\varphi}) \times \hat{\mathbf{z}}}{B_{0}} \cdot \nabla p_{0}=0,
\label{modifiedpressure}
\end{equation}
\begin{equation}
\eta J_{\|}=-\nabla_{\|}\left(\bar{\varphi}+\tilde{\varphi}\right).
\label{modifiedohm}
\end{equation}
Because the large-scale mode has a much longer time scale than the small-scale convective cells---i.e., there is a time-scale separation. We can use the method of averaging to separate the dynamics of different scales and derive the full set of equations for the model, which is listed as follows:
\begin{subequations} {\label{mainequation}}
\begin{align}
&\begin{aligned}[t]\bigg(\frac{\partial}{\partial t}&+\tilde{\boldsymbol{v}}\cdot\nabla\bigg) \nabla_{\perp}^{2} \bar{\varphi}=\\&-\frac{S}{\tau_{A}}\bigg[\nabla_{\|}^{(0)^{2}} \bar{\varphi}+\underbrace{\left(\nabla_{\perp} \cdot\langle\tilde{\boldsymbol{b}} \tilde{\boldsymbol{b}}\rangle\right) \cdot \nabla_{\perp} \bar{\varphi}}_{(a)}+\underbrace{\left\langle\nabla_{\|}^{(0)} \tilde{\boldsymbol{b}} \cdot \nabla_{\perp} \tilde{\varphi}\right\rangle}_{(b)}+\underbrace{\left\langle\left(\tilde{\boldsymbol{b}} \cdot \nabla_{\perp}\right) \nabla_{\|}^{(0)} \tilde{\varphi}\right\rangle}_{(c)}\bigg]-\frac{\kappa B_{0}}{\rho_{0}} \frac{\partial \bar{p}_{1}}{\partial y}, \end{aligned}\\
&\left(\frac{\partial}{\partial t}+\tilde{\boldsymbol{v}}\cdot\nabla\right) \bar{p}_{1}-\frac{\nabla \bar{\varphi} \times \hat{\mathbf{z}}}{B_{0}} \cdot \nabla p_{0}=0,\\
&\left(\frac{\partial}{\partial t}+\tilde{\boldsymbol{v}}\cdot\nabla\right) \nabla_{\perp}^{2} \tilde{\varphi}=-\frac{S}{\tau_{A}}\bigg[\nabla_{\|}^{(0)^{2}} \tilde{\varphi}+\underbrace{\left(\tilde{\boldsymbol{b}} \cdot \nabla_{\perp}\right) \nabla_{\|}^{(0)} \bar{\varphi}}_{(\alpha)}+\underbrace{\nabla_{\|}^{(0)}\left(\tilde{\boldsymbol{b}} \cdot \nabla_{\perp}\right) \bar{\varphi}}_{(\beta)}\bigg]-\frac{\kappa B_{0}}{\rho_{0}} \frac{\partial \tilde{p}_{1}}{\partial y},\\
&\left(\frac{\partial}{\partial t}+\tilde{\boldsymbol{v}}\cdot\nabla\right) \tilde{p}_{1}-\frac{\nabla \tilde{\varphi} \times \mathbf{z}}{B_{0}} \cdot \nabla p_{0}=0.
\end{align}
\end{subequations}
In Eq.(\ref{mainequation}), $\bar{\varphi}$ and $\bar{p}$ are the electrostatic potential and pressure of the large-scale mode, $\tilde{\varphi}$ and $\tilde{p}$ are electrostatic potential and pressure fluctuations of the small-scale convective cells, and $\boldsymbol{\tilde{v}}$ is the $\boldsymbol{E}\times\boldsymbol{B}$ velocity fluctuation due to $\tilde{\varphi}$. Since we are studying the dynamics of a single mode, the coupling between large-scale modes is not considered. $\rho_{0}$ is the plasma mass density, which is a constant.  $\boldsymbol{B_{0}}=B_{0}\boldsymbol{b_{0}}=B_{\phi}\boldsymbol{\hat{\phi}}+B_{\theta}(r)\boldsymbol{\hat{\theta}}$ $(B_{\phi}\gg B_{\theta})$ is the mean field, and $\boldsymbol{\tilde{b}}=\boldsymbol{\tilde{B}_{\bot}}/B_{0}$ is the direction of perturbed magnetic field. $\boldsymbol{B_{0}}$ and $\boldsymbol{\tilde{B}_{\bot}}$ together constitute the magnetic field configuration in this paper. $\boldsymbol{\kappa}=-\kappa\boldsymbol{\hat{r}}$ is the magnetic curvature, which has a dimension of length$^{-1}$. $p_{0}(r)$ is the mean pressure profile, which is the only source of the free energy in this model. Two characteristic time scales are adopted: $\tau_{A}=a(4\pi\rho_{0})^{1/2}/B_0$ is the Alfvén time, and
$\tau_R=4\pi a^2/\eta$ is the resistive diffusion time ($a$ is the characteristic width of the system, $\eta$ is the plasma resistivity). The ratio of $\tau_{R}$ to $\tau_{A}$ is denoted by $S$. Several different gradient operators are used in this model, and their definitions are as follows: $\nabla=\nabla_{\|}^{(0)}\boldsymbol{b_{0}}+\nabla_{\bot}$ is the gradient operator, $\nabla_{\bot}=\partial_{r}\boldsymbol{\hat{r}}+(\partial_{\theta}/r)\boldsymbol{\hat{\theta}}$ is the perpendicular gradient, $\nabla_{\|}^{(0)}=\boldsymbol{b_{0}}\cdot\nabla$ is the parallel gradient along the mean field $\boldsymbol{B_{0}}$, and $\nabla_{\|}=\nabla_{\|}^{(0)}+\boldsymbol{\tilde{b}}\cdot\nabla_{\bot}$ is the parallel gradient along the total field $\boldsymbol{B_{0}}+\boldsymbol{\tilde{B}_{\bot}}$. 

The bracket appearing in Eq.(\ref{mainequation}) is defined as the averaging over toroidal and poloidal directions, i.e., 
\begin{equation}
\langle A\rangle=\left(\frac{1}{2 \pi}\right)^{2} \iint d \theta d \phi e^{-i(m \theta-n \phi)} A,
\label{bracket}
\end{equation}
where $m$ and $n$ are the mode numbers of the low $m$, large-scale mode. After this averaging, only structures whose scales are comparable to that of the large-scale mode can be retained. Thus $\bar{\varphi}=\langle\varphi\rangle=\langle\left(\bar{\varphi}+\tilde{\varphi}\right)\rangle$, $\bar{p}_1=\langle p_1\rangle=\langle\left(\bar{p}_1+\tilde{p}_1\right)\rangle$.

The geometric configuration of plasma is taken to be a periodic cylinder.
\subsection{\label{classical}Resistive Interchange Mode in a Normal Magnetic Field}
In the absence of magnetic perturbation (no $\boldsymbol{\tilde{b}}$, no $\tilde{\varphi}$), Eq.(\ref{modifiedvorticity}), Eq.(\ref{modifiedpressure}), and Eq.(\ref{modifiedohm}) reduce to
\begin{equation}
\underbrace{\frac{\rho_{0}}{B_{0}^{2}}\frac{\partial}{\partial t} \nabla_{\perp}^{2} \varphi}_{-\nabla_{\bot}\cdot\boldsymbol{J_{pol}} }=\underbrace{\boldsymbol{b_{0}} \cdot \nabla J_{\|^{(0)}}}_{\nabla_{\|}^{(0)}J_{\|^{(0)}}}-\underbrace{\frac{\kappa}{B_{0}} \frac{\partial p_{1}}{\partial y}}_{-\nabla_{\bot}\cdot \boldsymbol{J_{PS}}},
\label{vorticity}
\end{equation}
\begin{equation}
\frac{\partial p_{1}}{\partial t}-\frac{\nabla \varphi \times \boldsymbol{{b}_{0}}}{B_{0}} \cdot \nabla p_{0}=0,\\
\label{pressure}
\end{equation}
\begin{equation}
\eta J_{\|^{(0)}}=-\boldsymbol{b_{0}} \cdot \nabla \varphi=-\nabla_{\|}^{(0)} \varphi.
\label{ohm}
\end{equation}
In Eq.(\ref{vorticity}), $\boldsymbol{J_{pol}}$ denotes the polarization current, and $\boldsymbol{J_{PS}}$ is the Phirsch–Schluter current. Eq.(\ref{vorticity}) is just the expanded form of $\nabla\cdot\boldsymbol{J}=0$, so quasi-neutrality is naturally maintained at lowest order.

The Fourier series for $\varphi$ and $p_1$ are
\begin{displaymath}
\begin{aligned}
&\varphi=\sum_{\boldsymbol{k}}\varphi_{\boldsymbol{k}}(x)e^{\gamma_{\boldsymbol{k}} t+i(m\theta-n\phi)},\\ &p_1=\sum_{\boldsymbol{k}}p_{1\boldsymbol{k}}\left(x\right)e^{\gamma_{\boldsymbol{k}} t+i(m\theta-n\phi)},
\end{aligned}
\end{displaymath}
where $\boldsymbol{k}$ denotes a set of mode number $(m,n)$ and $x=r-r_{mn}$ is a coordinate denoting the distance from the position of the resonant surface $r_{mn}$. Plugging in these series, Eq.(\ref{vorticity}), Eq.(\ref{pressure}), and Eq.(\ref{ohm}) then reduce to the following eigenmode equation (~\cite{FKR,Coppi1966,Carreras&Garcia&PD1993})
\begin{equation}
-\gamma_{\boldsymbol{k}} \frac{\partial^{2} \varphi_{\boldsymbol{k}}}{\partial x^{2}}+\frac{S}{\tau_{A}} \frac{k_{\theta}^{2}}{L_{S}^{2}} x^{2} \varphi_{\boldsymbol{k}}+\left(\gamma_{\boldsymbol{k}} k_{\theta}^{2}-\frac{\kappa p_{0} k_{\theta}^{2}}{\rho_{0} L_{p} \gamma_{\boldsymbol{k}}}\right) \varphi_{\boldsymbol{k}}=0,
\label{eigen}
\end{equation}
where
\begin{displaymath}
\begin{aligned}
L_{p}&=\left|(1/p_{0})(\textrm{d}p_{0}/\textrm{d}r)\right|^{-1}& \quad L_{s}&=s/Rq,\\ 
s&=\left|\textrm{dln}(q)/\textrm{dln}(r)\right|& \quad k_{\theta}&=m/r_{mn},
\end{aligned}
\end{displaymath}
and $R$ is the major radius of the torus.

Eq.(\ref{eigen}) can be solved exactly, and its eigen solutions are listed in Appendix.(\ref{s&f}). In this paper, we merely use the growth rate of ``ground state" in two limiting cases:
\begin{itemize}
\item $k_{r}\gg k_{\theta}$ (slow interchange ordering)
\begin{equation}
\gamma_{\boldsymbol{k}}=S^{-\frac{1}{3}}\tau_{A}^{\frac{1}{3}} \tau_{p}^{-\frac{2}{3}} \tau_{\kappa}^{-\frac{2}{3}} \tilde{k}_{\theta}^{\frac{2}{3}};
\label{slow}
\end{equation}
\item $k_{r}\ll k_{\theta}$ (fast interchange ordering)
\begin{equation}
\gamma_{\boldsymbol{k}}=\tau_{p}^{-\frac{1}{2}} \tau_{\kappa}^{-\frac{1}{2}},
\label{fast}
\end{equation}
\end{itemize}
where
\begin{displaymath}
\tau_{p}=L_{p}/c_{s}\quad  \tau_{\kappa}=1/c_{s}\kappa\quad \tilde{k}_{\theta}=k_{\theta}L_{s}.
\end{displaymath}
Detailed calculations of the above results can also be find in Appendix.(\ref{s&f}).
\subsection{\label{introstoch}Model with a Static Stochastic Magnetic Field}
In this section, we discuss the effects of the stochastic magnetic field.

\begin{figure}[h]
\includegraphics[scale=0.5]{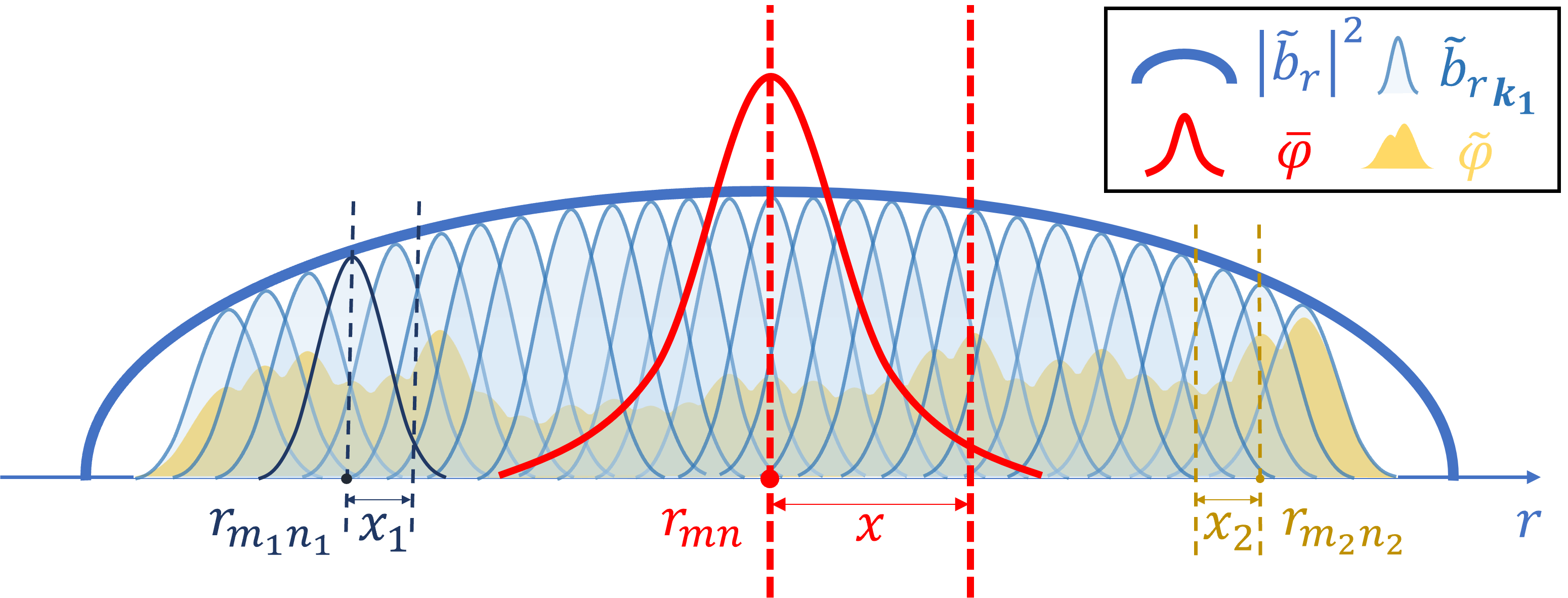}
\caption{\label{configuration} Basic configuration of the model: thick blue line represents the profile of stochastic magnetic field and small humps are perturbed magnetic fields at different resonant surfaces; red line represents the large-scale mode; yellow lump is the resultant microturbulence.}
\end{figure}

With RMP, magnetic field lines become chaotic in the edge layer. So the total magnetic field consists of a main field and a perturbed field, i.e., $\boldsymbol{b_{tot}}=\boldsymbol{b_0}+\boldsymbol{\tilde{b}}$. More specifically, $\boldsymbol{\tilde{b}}$ is composed of a series of small magnetic perturbations strongly localized at resonant surfaces, as shown in FIG.~\ref{configuration}. When these $\tilde{b}_{r_{\boldsymbol{k_{1}}}}$ are sufficiently densely packed so magnetic islands overlap, the field becomes stochastic. Therefore, the way we introduce stochastic magnetic field is by modifying the parallel gradient $\nabla_{\|}^{(0)}$ to $\nabla_{\|}$, which refers to $\nabla_{\|}\rightarrow\nabla_{\|}^{(0)}+\boldsymbol{\tilde{b}}\cdot\nabla_{\bot}$.

\begin{figure}[h]
    \centering
    \includegraphics[scale=0.6]{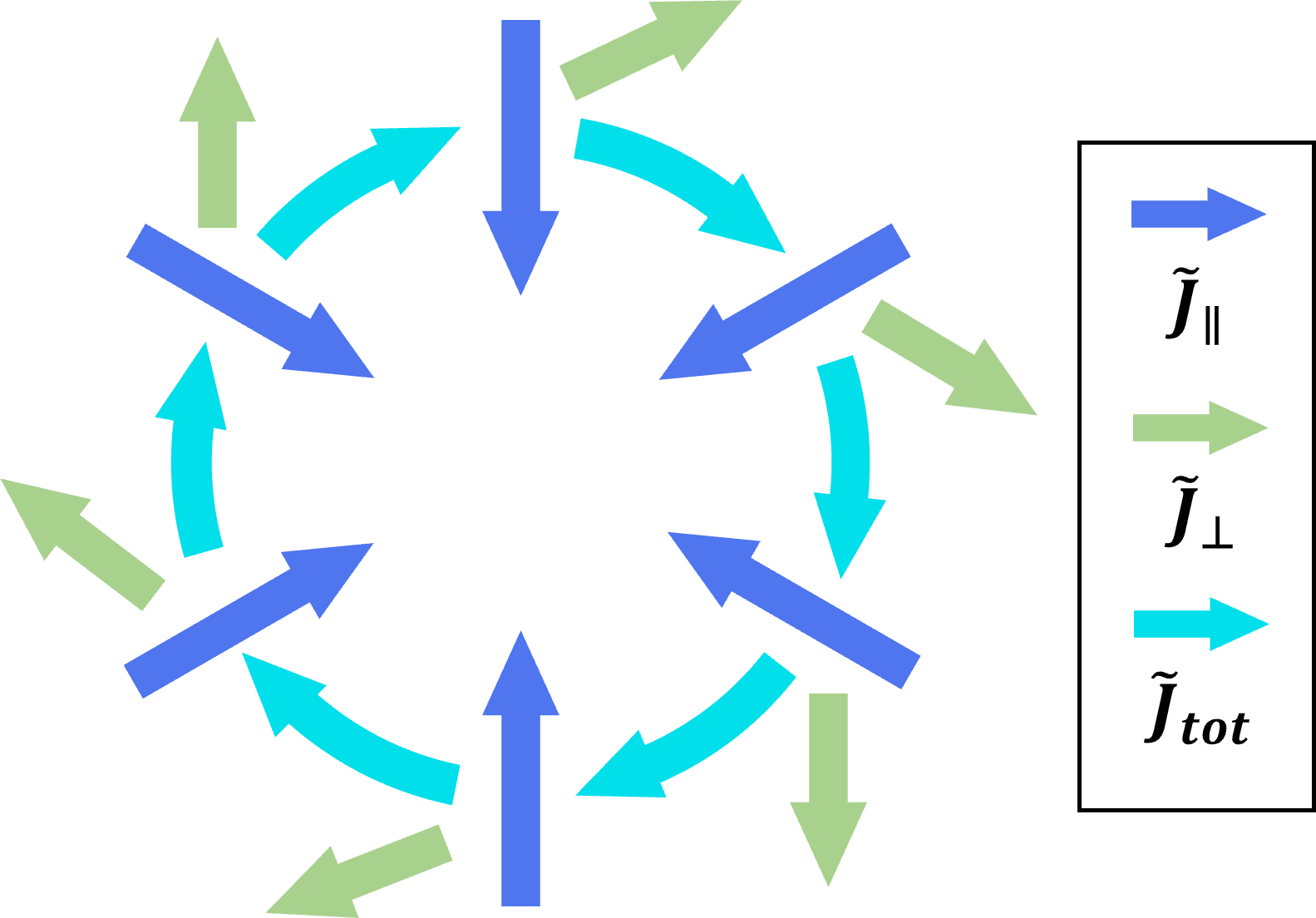}
    \caption{The balance between $\boldsymbol{\tilde{J}_{\|}}$ and $\boldsymbol{\tilde{J}_{\bot}}$. The current density fluctuation along perturbed field lines $\boldsymbol{\tilde{J}_{\bot}}$ is not itself divergence free. An extra current density fluctuation $\boldsymbol{\tilde{J}_{\bot}}$ is able to balance $\boldsymbol{\tilde{J}_{\|}}$ so that the total current density fluctuation $\boldsymbol{\tilde{J}_{tot}}$ is divergence free.}
    \label{currentbalance}
\end{figure}

With this change, the parallel current density given by Eq.(\ref{ohm}) is modified to
\begin{equation}
\boldsymbol{J}_{\|}=-\frac{1}{\eta}\left[\nabla_{\|}^{(0)}+\tilde{\boldsymbol{b}}\cdot\nabla_{\bot}\right]\bar{\varphi}(\boldsymbol{b_{0}}+\tilde{\boldsymbol{b}}),
\label{paracurrent}
\end{equation}
N.B. here $\boldsymbol{J_{\|}}$ refers to current along the wandering field. This will necessarily render $\nabla\cdot\boldsymbol{J}=0$ a stochastic differential equation. Since the scale of the single mode is much larger than that of the stochastic magnetic field, Eq.(\ref{paracurrent}) indicates that the stochastic magnetic field can induce a small-scale current $\boldsymbol{\tilde{J}_{\|}}$, and its divergence is
\begin{equation}
\widetilde{\nabla_{\|}\boldsymbol{J}_{\|}}=-\frac{1}{\eta}\bigg[\underbrace{\left(\tilde{\boldsymbol{b}}\cdot\nabla_{\perp}\right)\nabla_{\|}^{(0)}\bar{\varphi}}_{(\alpha)}+\underbrace{\nabla_{\|}^{(0)}\left(\tilde{\boldsymbol{b}}\cdot\nabla_{\perp}\right)\bar{\varphi}}_{(\beta)}\bigg].
\label{div}
\end{equation}
The quasi-neutrality condition requires $\nabla\cdot\boldsymbol{J}=0$ at all scales, which means $\boldsymbol{\tilde{J}_{\|}}$ should also be divergence-free. To clarify, we need to use the Fourier series of $\boldsymbol{\tilde{b}}$ and $\bar{\varphi}$:
\begin{displaymath}
\begin{aligned}
\tilde{\boldsymbol{b}}&=\sum_{\boldsymbol{k_{1}}}\tilde{\boldsymbol{b}}_{\boldsymbol{k}_{1}}\left(x_{1}\right) e^{i\left(m_{1} \theta-n_{1} \phi\right)},\\
\bar{\varphi}&=\bar{\varphi}_{\boldsymbol{k}}(x) e^{\gamma_{\boldsymbol{k}}t+i(m \theta-n \phi)},
\end{aligned}
\end{displaymath}
where
$x_1=r-r_{m_{1}n_{1}}$, $x=r-r_{mn}$, as illustrated in FIG.~\ref{configuration}. Plugging these Fourier series into Eq.(\ref{div}), terms labelled as $(\alpha)$ and $(\beta)$ are then equal to
\begin{equation}
\begin{aligned}
&\left(\tilde{\boldsymbol{b}} \cdot \nabla_{\perp}\right) \nabla_{\|}^{(0)} \bar{\varphi} \\
=& \sum_{\boldsymbol{k_{1}}}\left[
\tilde{b}_{r_{\boldsymbol{k_{1}}}}\left(x_{1}\right) e^{i\left(m_{1} \theta-n_{1} \phi\right)} \partial_{x}\left(i k_{\|} \bar{\varphi}_{\boldsymbol{k}}(x) e^{i(m \theta-n \phi)}\right)
\right]+\\
&\sum_{\boldsymbol{k_{1}}}\left[
+\tilde{b}_{\theta_{\boldsymbol{k_{1}}}}\left(x_{1}\right) e^{i\left(m_{1} \theta-n_{1} \phi\right)} \partial_{y}\left(i k_{\|} \bar{\varphi}_{\boldsymbol{k}}(x) e^{i(m \theta-n \phi)}\right)
\right]\\
\approx & \sum_{\boldsymbol{k_{1}}} \tilde{b}_{r_{\boldsymbol{k_{1}}}}\left(x_{1}\right) \partial_{x}\left(i k_{\|} \bar{\varphi}_{\boldsymbol{k}}(x)\right) e^{i\left[\left(m_{1}+m\right) \theta-\left(n_{1}+n\right) \phi\right]},
\end{aligned}
\label{alpha}
\end{equation}
and
\begin{equation}
\begin{aligned}
& \nabla_{\|}^{(0)}\left(\tilde{\boldsymbol{b}} \cdot \nabla_{\perp}\right) \bar{\varphi} \\
=& \nabla_{\|}^{(0)}\sum_{\boldsymbol{k_{1}}}\left[
\tilde{b}_{r_{\boldsymbol{k_{1}}}}\left(x_{1}\right) e^{i\left(m_{1} \theta-n_{1} \phi\right)} \partial_{x} \left(\bar{\varphi}_{\boldsymbol{k}}(x) e^{i(m \theta-n \phi)}\right)
\right]+\\
& \nabla_{\|}^{(0)}\sum_{\boldsymbol{k_{1}}}\left[
\tilde{b}_{\theta_{\boldsymbol{k_{1}}}}\left(x_{1}\right) e^{i\left(m_{1} \theta-n_{1} \phi\right)} \partial_{y} \left(\bar{\varphi}_{\boldsymbol{k}}(x) e^{i(m \theta-n \phi)}\right)
\right]\\
\approx& \sum_{\boldsymbol{k}_{\mathbf{1}}} i\left(k_{1 \|}+k_{\|}\right) \tilde{b}_{r_{\boldsymbol{k}_{\mathbf{1}}}}\left(x_{1}\right) \partial_{x} \bar{\varphi}_{\boldsymbol{k}}(x) e^{i\left[\left(m_{1}+m\right) \theta-\left(n_{1}+n\right) \phi\right]},
\end{aligned}
\label{beta}
\end{equation}
in which the slow interchange ordering approximation has been used. In order for $\widetilde{\nabla_{\|}J_{\|}}=0$ to be true, the equation
\begin{equation}
\left[k_{1\theta}\left(x+r_{mn}-r_{m_{1}n_{1}}\right)\partial_{x}\bar{\varphi}+k_{\theta}\bar{\varphi}\right]=0
\label{necessary}
\end{equation}
must be true for arbitrary $\boldsymbol{k_{1}}$, which is clearly impossible. Thus to maintain $\nabla\cdot\boldsymbol{J}=0$, a small-scale electrostatic potential fluctuation $\tilde{\varphi}$ and a small-scale pressure fluctuation $\tilde{p}_1$ must be driven, which can make extra contributions to $\boldsymbol{\tilde{J}_{\bot}}$ so as to keep $\widetilde{\nabla_{\|}J_{\|}}+\widetilde{\nabla_{\bot}\cdot\boldsymbol{{J}_{\bot}}}=0$, as illustrated in FIG.~\ref{currentbalance}.

\begin{figure}[h]
    \centering
    \includegraphics[scale=0.5]{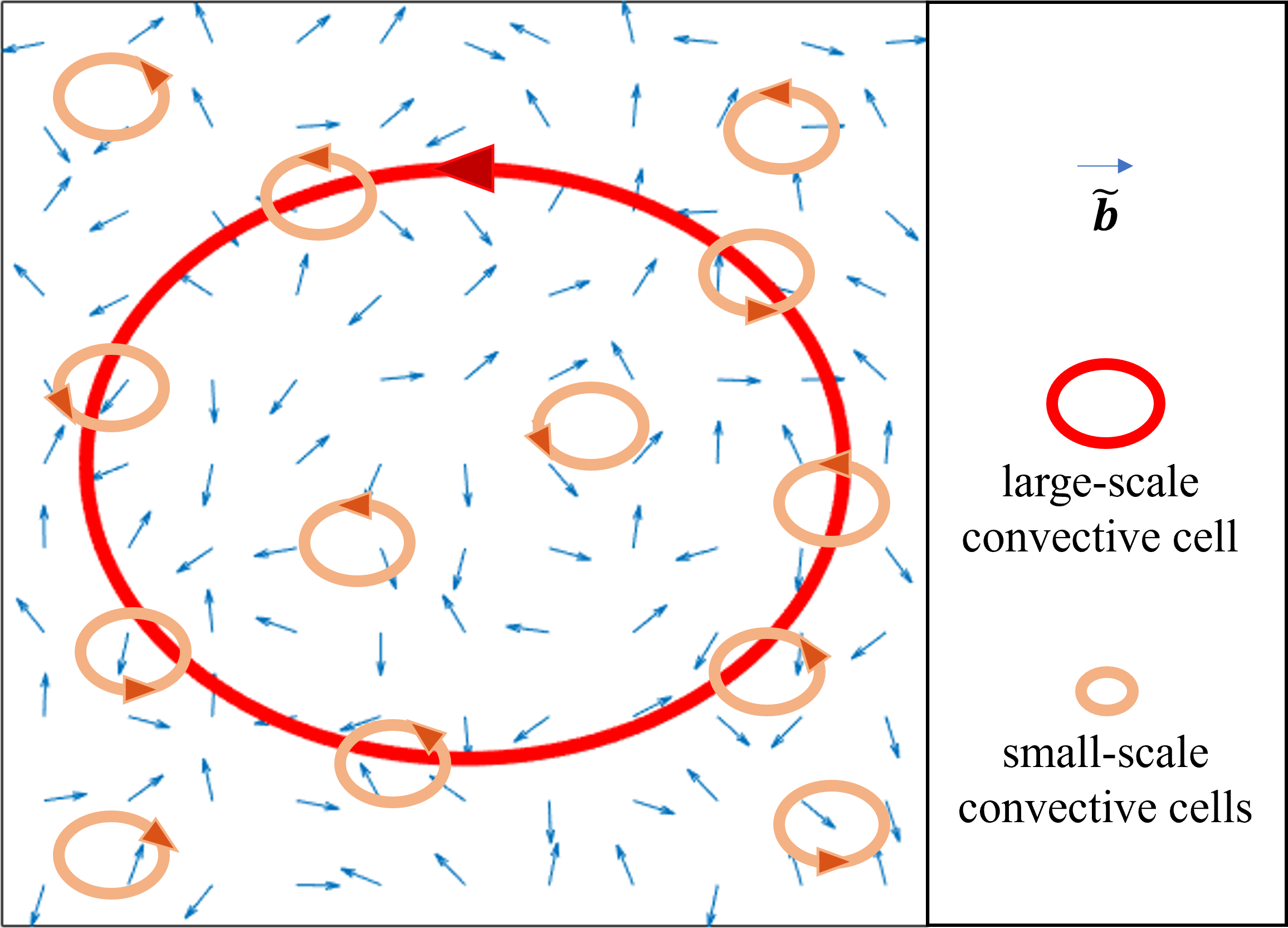}
    \caption{Illustration of three main players in this model: large-scale single cell (large red ellipse), prescribed static stochastic magnetic field background (small blue arrows), and small-scale convective cells (small orange ellipsis). As the large-scale mode and small-scale convective cells interact, this is a multi-scale problem.}
    \label{player}
\end{figure}

Physically, the appearance of $\tilde{\varphi}$ signals the presence of small-scale convective cells. Therefore, this system actually contains three players: a large-scale single cell, a prescribed stochastic magnetic field background, and small-scale convective cells, i.e., the intrinsic multi-scale microturbulence (see FIG.~\ref{player}). These convective cells are driven by the beating of the large-scale mode and the stochastic field. This fact brings us back to Eq.(\ref{modifiedvorticity}), Eq.(\ref{modifiedpressure}) and Eq.(\ref{modifiedohm}), which further yield the full set of model equations Eq.(\ref{mainequation}a)$\sim$ Eq.(\ref{mainequation}d), by using method of averaging. This idea is similar to that of Kadomtsev and Pogutse's study in 1979, in which small-scale temperature fluctuations are generated to maintain $\nabla\cdot\boldsymbol{q}=0$ ($\boldsymbol{q}$ is electron heat flux) at all scales in a stochastic magnetic field~\cite{K&P1979}. See Appendix.(\ref{KP}) for more details. 

For simplicity, the fast interchange ordering approximation is applied to $\tilde{\varphi}$, as it is small scale. Thus the fundamental scale ordering is
\begin{equation}
\underbrace{\frac{1}{k_{1\theta}}\ll w_{\boldsymbol{k_{1}}}}_{fast}\ll \underbrace{w_{\boldsymbol{k}}\ll\frac{1}{k_{\theta}}}_{slow},
\label{scaleordering}
\end{equation}
where $w_{\boldsymbol{k_{1}}}$ and $w_{\boldsymbol{k}}$ are characteristic radial widths of $\bar{\varphi}_{\boldsymbol{k}}$ and $\tilde{\varphi}_{\boldsymbol{k_{1}}}$, respectively. 

Moreover, this convective cell microturbulence will generate a turbulent viscosity $\nu$ and a turbulent diffusivity $\chi$. Due to the separation of temporal scales, and since the large-scale cell is evolving very slowly, the small-scale cells are considered in a stationary state, saturated by $\nu$ and $\chi$. In this light, we can approximate the nonlinear operator $\boldsymbol{{\tilde{v}}}\cdot\nabla$ in Eq.(\ref{mainequation}) by a renormalized diffusion operator $-\nu\nabla_{\bot}^{2}$. We further take $\nu$ and $\chi$ as equal, as their physical mechanisms are both random advection. As will be seen in Sec.~\ref{turbulentvis}, the scaling of $\nu$ is calculated through nonlinear closure theory. 

Obviously, Eq.(\ref{mainequation}a) and Eq.(\ref{mainequation}c) are coupled to each other, since implies large and small scale dynamics are connected: the beat of $\tilde{\boldsymbol{b}}$ and $\bar{\varphi}$ serves as the drive for $\tilde{\varphi}$, while $\tilde{\varphi}$ modifies the evolution of $\bar{\varphi}$. This relation is illustrated in the feedback loop FIG.~\ref{flowchart}.

\subsection{\label{tildevarphi}Response of $\tilde{\varphi}$ to $\tilde{\boldsymbol{b}}$}
To calculate the growth rate of $\bar{\varphi}$, we need to deal with the correlations on the R.H.S. of Eq.(\ref{mainequation}a). This requires us to find the relation between $\tilde{\varphi}$ and $\tilde{\boldsymbol{b}}$, which is implicit in Eq.(\ref{mainequation}c). We can use Eq.(\ref{mainequation}d) to eliminate $\tilde{p}_1$ in Eq.(\ref{mainequation}c). The Fourier series of $\tilde{p}$ and $\tilde{\varphi}$ are
\begin{displaymath}
\begin{aligned}
\tilde{p}&=\sum_{\boldsymbol{k_{2}}}\tilde{p}_{\boldsymbol{k_{2}}}(x_{2})e^{\gamma_{\boldsymbol{k}}t+i(m_{2}\theta-n_{2}\phi)},\\
\tilde{\varphi}&=\sum_{\boldsymbol{k_{2}}}\tilde{\varphi}_{\boldsymbol{k_{2}}}(x_{2})e^{\gamma_{\boldsymbol{k}}t+i(m_{2}\theta-n_{2}\phi)},\\
\end{aligned}
\end{displaymath}
where
\begin{displaymath}
x_{2}=r-r_{m_{2}n_{2}}.
\end{displaymath}
N.B. the growth rates of $\tilde{p}_{\boldsymbol{k_{2}}}$ and $\tilde{\varphi}_{\boldsymbol{k_{2}}}$ are $\gamma_{\boldsymbol{k}}$, instead of $\gamma_{\boldsymbol{k_{2}}}$. The growth of $\tilde{\varphi}$ and $\tilde{p}$ can be viewed at two different time scales. On a short time scale ($\sim1/\gamma_{\boldsymbol{k_{2}}}$), $\tilde{\varphi}$ and $\tilde{p}$ are driven by the magnetic curvature and pressure gradient, and damped by $\nu$ and $\chi$. Because of the damping of $\nu$ and $\chi$, $\tilde{\varphi}$ and $\tilde{p}$ can relax to a stationary state at this time scale. But on a much longer time scale ($\sim1/\gamma_{\boldsymbol{k}}$), because large scale and small scales interact, the magnitudes of $\tilde{p}$ and $\tilde{\varphi}$ are adiabatically modulated by the magnitude of $\bar{\varphi}$. Therefore, even though fast interchange approximation is applied to $\tilde{\varphi}$ and $\tilde{p}$, their actual growth rates are the same as those of $\bar{\varphi}$ and $\bar{p}$, due to the turbulent viscous and thermal diffusive stresses induced by $\nu$ and $\chi$. Then plugging these series into Eq.(\ref{mainequation}d) and utilizing the approximation $\gamma_{\boldsymbol{k}}\ll\nu k_{2\theta}$, we get
\begin{equation}
\tilde{p}_{1 \boldsymbol{k_{2}}}=-\frac{i p_{0}}{\chi k_{2 \theta} B_{0} L_{p}} \tilde{\varphi}_{\boldsymbol{k_{2}}}.
\label{ptophi}
\end{equation}
By combining Eq.(\ref{alpha}), Eq.(\ref{beta}), Eq.(\ref{mainequation}c), and Eq.(\ref{ptophi}), we obtain an equation whose L.H.S. is homogeneous in $\tilde{\varphi}$, and whose R.H.S is the $\tilde{\boldsymbol{b}}\bar{\varphi}$ drive:
\begin{equation}
\begin{aligned}
&\left[-2 \nu k_{2 \theta}^{2} \frac{\partial^{2}}{\partial x_{2}^{2}}+\frac{S}{\tau_{A}} \frac{k_{2 \theta}^{2} x_{2}^{2}}{L_{s}^{2}}-\left(\frac{1}{\chi \tau_{p}\tau_{\kappa}}-\nu k_{2 \theta}^{4}\right)\right] \tilde{\varphi}_{\boldsymbol{k_{2}}}\left(x_{2}\right) \\
&= i \frac{S}{\tau_{A}}\left[\left(\partial_{x} k_{\|}\right) \bar{\varphi}_{\boldsymbol{k}}(x)+\left(k_{2 \|}+k_{\|}\right) \partial_{x} \bar{\varphi}_{\boldsymbol{k}}(x)\right] \tilde{b}_{r_{\boldsymbol{{\left(k_{2}-k\right)}}}},
\end{aligned}
\label{microdrive}
\end{equation}
The operator on the L.H.S. of Eq.(\ref{microdrive}) looks like that for a quantum harmonic oscillator. So we define the following quantities:
\begin{displaymath}
M_{\boldsymbol{k_{2}}}=\frac{1}{4 \nu k_{2 \theta}^{2}}, \Omega_{\boldsymbol{k_{2}}}=\sqrt{\frac{8 \nu S k_{2 \theta}^{4}}{\tau_{A} L_{s}^{2}}}, \Lambda_{\boldsymbol{k_{2}}}=\frac{1}{\chi \tau_{p}\tau_{\kappa}}-\nu k_{2 \theta}^{4}.
\end{displaymath}
Then the corresponding Green's function for Eq.(\ref{microdrive}) is
\begin{equation}
G\left(x_{2}, x_{2}^{\prime}\right)=\sum_{l} \frac{\psi_{\boldsymbol{k_{2}}}^{l}\left(x_{2}\right) \psi_{\boldsymbol{k_{2}}}^{l}\left(x_{2}^{\prime}\right)}{\Lambda_{\boldsymbol{k_{2}}}^{l}-\Lambda_{\boldsymbol{k_{2}}}},
\label{green}
\end{equation}
where
\begin{displaymath}
\begin{aligned}
 &\psi_{\boldsymbol{{k_{2}}}}^{l}\left(x_{2}\right)=\frac{w_{\boldsymbol{{k_{2}}}}^{1 / 2}}{\pi^{1 / 4}} \frac{1}{\sqrt{2^{l} l !}} H_{l}\left(w_{\boldsymbol{{k_{2}}}} x_{2}\right) e^{-\frac{\left(w_{\boldsymbol{{k_{2}}}} x_{2}\right)^{2}}{2}},\\ w_{\boldsymbol{{k_{2}}}}&=\left(\frac{S}{2 \tau_{A} \nu L_{s}^{2}}\right)^{1 / 4}, \quad\Lambda_{\boldsymbol{{k_{2}}}}^{l}=\sqrt{\frac{8 \nu S k_{2 \theta}^{4}}{\tau_{A} L_{s}^{2}}}\left(l+\frac{1}{2}\right).
\end{aligned}
\end{displaymath}
By making use of this Green's function, the solution to Eq.(\ref{microdrive}) is 
\begin{equation}
\begin{aligned}
\tilde{\varphi}_{\boldsymbol{k_{2}}}&=-i\frac{S}{\tau_{A}}\times\\
& \int\left[\partial_{x^{\prime}}\left(Gk_{2 \|} \tilde{b}_{r_{\left(\boldsymbol{k_{2}}-\boldsymbol{k}\right)}}\right)+k_{\|} \partial_{x^{\prime}}\left( G\tilde{b}_{r_{\left(\boldsymbol{k_{2}}-\boldsymbol{k}\right)}}\right)\right] \bar{\varphi}_{\boldsymbol{k}} d x_{2}^{\prime}
\end{aligned}
\label{response0}
\end{equation}
As ${\bar{\varphi}}_{\boldsymbol{k}}$ varies much more slowly than ${\tilde{b}}_{r_{\left(\boldsymbol{k_{2}}-\boldsymbol{k}\right)}}$ (see Fig.1), it is reasonable to move ${\bar{\varphi}}_{\boldsymbol{k}}$ out of the integral in Eq.(\ref{response0}), and approximate it by its value at $x=0$. Then the first term of the integrand vanishes, because it is a total derivative. Using integration by parts, the response of $\tilde{\varphi}_{\boldsymbol{k_2}}$ to $\tilde{b}_{r_{\boldsymbol{(k_{2}-k)}}}$ is then approximately
\begin{equation}
\tilde{\varphi}_{\boldsymbol{k_{2}}} \approx- i\frac{k_{\theta}}{L_{s}} \frac{S}{\tau_{A}} \sum_{l} \frac{\psi_{\boldsymbol{k_{2}}}^{l}\left(x_{2}\right)}{\Lambda_{\boldsymbol{k_{2}}}^{l}-\Lambda_{\boldsymbol{k}_{2}}} \bar{\varphi}_{\boldsymbol{k}}(0) \int \psi_{\boldsymbol{k_{2}}}^{l} \tilde{b}_{r_{\left(\boldsymbol{k_{2}}-\boldsymbol{k}\right)}} d x_{2}^{\prime}.
\label{response}
\end{equation}

\subsection{\label{finalresult}Corrected Growth Rate and Scaling of Turbulent Viscosity}
Utilizing the result of Eq.(\ref{response}), we can simplify the correlations which appear in Eq.(\ref{mainequation}a), and calculate the correction to the single mode's growth rate. As will be seen, the turbulent viscosity $\nu$ still appears in the expression for the growth rate as an unknown quantity. It can be calculated by using closure theory.
\subsubsection{\label{correlations}Correlation $(a)$, $(b)$, and $(c)$}
The three correlations we need to calculate are
\begin{equation}
\begin{aligned}
(a)&=\left(\nabla_{\perp} \cdot\langle\tilde{\boldsymbol{b}} \tilde{\boldsymbol{b}}\rangle\right) \cdot \nabla_{\perp} \bar{\varphi},\\
(b)&=\left\langle\nabla_{\|}^{(0)} \tilde{\boldsymbol{b}} \cdot \nabla_{\perp} \tilde{\varphi}\right\rangle,\\
(c)&=\left\langle\left(\tilde{\boldsymbol{b}} \cdot \nabla_{\perp}\right) \nabla_{\|}^{(0)} \tilde{\varphi}\right\rangle.
\end{aligned}
\end{equation}

Consistent with the slow interchange ordering at large scale, $\tilde{b}_{\theta}\partial_{y}\bar{\varphi}\ll\tilde{b}_{r}\partial_{r}\bar{\varphi}$, and $\boldsymbol{\tilde{b}}\cdot\nabla_{\perp}\bar{\varphi}\approx b_{r}\partial_{r}\bar{\varphi}$. Therefore, correlation $(a)$ can be rewritten as
\begin{equation}
(a)=\left(i k_{\theta}\right)\left|\tilde{b}_{\theta} \tilde{b}_{r}\right| \partial_{x} \bar{\varphi}+\partial_{x}\left(\left|\tilde{b}_{r}^{2}\right| \partial_{x} \bar{\varphi}\right)\approx\left|\tilde{b}_{r}^{2}\right| \partial_{x}^{2} \bar{\varphi},
\label{correlation(a)}
\end{equation}
as $1/k_{\theta}$ is largest scale of the ordering in Eq.(\ref{scaleordering}), and $\left|\tilde{b}_{r}^{2}\right|$ varies more slowly than $\bar{\varphi}$ does.

For correlation $(b)$, 
\begin{equation}
\begin{aligned}
(b)=&\sum_{\boldsymbol{k_{2}}} i k_{\|} \left(i k_{\theta}\right) \tilde{b}_{\theta_{\boldsymbol{k}-\boldsymbol{k}_{2}}} \tilde{\varphi}_{\boldsymbol{{k_{2}}}}+\sum_{\boldsymbol{{k_{2}}}} i k_{\|} \partial_{x}\left[\tilde{b}_{r_{\boldsymbol{k}-\boldsymbol{k}_{2}}} \tilde{\varphi}_{\boldsymbol{k}_{2}}\right]\\
\approx&\sum_{\boldsymbol{k_{2}}} i k_{\|} \partial_{x}\left[\tilde{b}_{r_{\boldsymbol{k}-\boldsymbol{k}_{2}}}\left(x_{1}\right) \tilde{\varphi}_{\boldsymbol{k_{2}}}\left(x_{2}\right)\right],
\end{aligned}
\label{correlation(b)}
\end{equation}

Similarly, for correlation $(c)$,
\begin{equation}
\begin{aligned}
(c)=&\sum_{\boldsymbol{k_{2}}} \partial_{x}\left[i k_{2 \|} \tilde{b}_{r_{\boldsymbol{k}-\boldsymbol{k}_{2}}} \tilde{\varphi}_{\boldsymbol{k}_{2}}\right]+\sum_{\boldsymbol{k}_{2}} \left(i k_{\theta}\right) i k_{2 \|} \tilde{b}_{\theta_{\boldsymbol{k}_{1}}} \tilde{\varphi}_{\boldsymbol{k}_{2}}\\
\approx&\sum_{\boldsymbol{k_{2}}} \partial_{x}\left[i k_{2 \|} \tilde{b}_{r_{\boldsymbol{k}-\boldsymbol{k}_{2}}} \tilde{\varphi}_{\boldsymbol{k}_{2}}\right].
\end{aligned}
\label{correlation(c)}
\end{equation}
By using the dense packing approximation (see FIG.~\ref{configuration}), the summation over $\boldsymbol{k_2}$ can be replaced by an integral. More specifically,
\begin{equation}
\sum_{\boldsymbol{k_{2}}\left(m_{2}, n_{2}\right)}=\iint d m_{2} d n_{2}=\frac{R}{L_{s}} \int d k_{2 \theta}\left|k_{2 \theta}\right| \int r d x.
\label{sumtointegral}
\end{equation}
Because Eq.(\ref{correlation(c)}) is a total derivative, it vanishes after writing the sum as an integral. Therefore, the correlations $(a)$, $(b)$, and $(c)$ are:
\begin{equation}
\begin{aligned}
&(a)\approx\left|\tilde{b}_{r}^{2}\right| \partial_{x}^{2} \bar{\varphi},\\
&(b)\approx-\frac{S}{\tau_{A}} \frac{R k_{\theta}^{2}}{L_{s}^{3}} \bar{\varphi}_{\boldsymbol{k}}(0)\left(x+r_{m n}\right)\times I,\\
&(c)\approx0
\end{aligned}
\label{correlaitonlist}
\end{equation}
where
\begin{displaymath}
I=\int d k_{2 \theta} \frac{\left|k_{2 \theta}\right|}{\Lambda_{k_{2}}^{0}-\Lambda_{k_{2}}}\left[\int d x_{2}^{\prime} b_{r_{k-k_{2}}} \psi_{k_{2}}^{0}\right]^{2}.
\end{displaymath}
Here, only the first term of the Green's function in Eq.(\ref{green}) is kept, since the magnitude of $\psi_{\boldsymbol{k_{2}}}^{l}$ decreases exponentially with $l$. If the spatial shape of each $\tilde{b}_{r_{\boldsymbol{k_2}}}(x_{2})$ is approximately Gaussian, then $I$ can be simplified to an integral over $k_{2\theta}$ which is
\begin{displaymath}
\int d k_{2 \theta}\left|k_{2 \theta}\right| \frac{\pi^{\frac{1}{2}}c^{2} Z^{2}\left(k_{\theta}-k_{2 \theta}\right) w_{\boldsymbol{k_{2}}} }{\Lambda_{\boldsymbol{k_{2}}}^{0}-\Lambda_{\boldsymbol{k_{2}}}} \left(\frac{1}{o_{\boldsymbol{k_{2}}}^{2}}+\frac{w_{\boldsymbol{k_{2}}}^{2}}{2}\right)^{-1},
\end{displaymath}
where $c$, $Z(k_{2\theta})$ and $o_{\boldsymbol{k_{2}}}$ are a normalization factor, spectrum and the characteristic width of $\tilde{b}_{r_{\boldsymbol{k_{2}}}}$ respectively. Because the width of $\tilde{b}_{\boldsymbol{k_{2}}}$ is much smaller than the width of $\tilde{\varphi}_{\boldsymbol{k_{2}}}$, i.e., $o_{\boldsymbol{k_{2}}}\ll 1/w_{\boldsymbol{k_{2}}}$, we can approximate $I$ as
\begin{displaymath}
\int d k_{2 \theta}\left|k_{2 \theta}\right| \frac{\pi^{\frac{1}{2}}c^{2} Z^{2}\left(k_{\theta}-k_{2 \theta}\right) w_{\boldsymbol{k_{2}}} o^{2}_{\boldsymbol{k_{2}}}}{\Lambda_{\boldsymbol{k_{2}}}^{0}-\Lambda_{\boldsymbol{k_{2}}}}.
\end{displaymath}
Since we have obtained the linear response of $\tilde{\varphi}_{\boldsymbol{k_{2}}}$ to $\tilde{b}_{r_{\left(\boldsymbol{k-k_{2}}\right)}}$ in Sec.~\ref{tildevarphi}, the correlation of $\tilde{b}_{r}$ and $\tilde{v}_{r}$ then is
\begin{equation}
\begin{aligned}
\left\langle\tilde{b}_{r} \tilde{v}_{r}\right\rangle=&\pi^{\frac{1}{2}} \frac{\tilde{k}_{\theta} R r_{m n} }{ L_{s}^{3}B_{0}} \frac{S}{\tau_{A}}\bar{\varphi}_{\boldsymbol{k}}(0)\times\\ 
&\int d k_{2 \theta} \left|k_{2 \theta}\right| k_{2 \theta} \frac{c^{2} Z^{2}\left(k_{\theta}-k_{2 \theta}\right) w_{\boldsymbol{k}_{2}} o_{\boldsymbol{k}_{2}}^{2}}{\Lambda_{\boldsymbol{k}_{\mathbf{2}}}^{0}-\Lambda_{\boldsymbol{k}_{\mathbf{2}}}}.
\label{bvcorrelation}
\end{aligned}
\end{equation}
Eq.(\ref{bvcorrelation}) indicates that $\langle\tilde{b}_{r}\tilde{v}_{r}\rangle$ is non-trivial in this model. \emph{$\langle\tilde{b}_{r}\tilde{v}_{r}\rangle\neq 0$ means the electrostatic turbulence phase locks to the ambient magnetic perturbations.} This is a direct result of Eq.(\ref{microdrive}), because $\tilde{b}_{r_{\left(\boldsymbol{k-k_{2}}\right)}}$ is the drive of $\tilde{\varphi}_{\boldsymbol{k_{2}}}$. Thus, the statistics of $\tilde{b}_{r}$ and $\tilde{v}_{r}$ are not independent.

\subsubsection{\label{finalcal}Corrected Growth Rate of $\bar{\varphi}$}
Substituting Eq.(\ref{correlaitonlist}) into Eq.(\ref{mainequation}a), and taking the Fourier transform of Eq.(\ref{mainequation}a) and Eq.(\ref{mainequation}b), we get
\begin{equation}
\begin{aligned}
&-\frac{S}{\tau_{A}} \frac{k_{\theta}^{2}}{L_{s}^{2}} \frac{d^{2}}{d k_{x}^{2}} \hat{\bar{\varphi}}_{\boldsymbol{k}}\left(k_{x}\right)+\gamma_{\boldsymbol{k}} k_{x}^{2} \hat{\bar{\varphi}}\left(k_{x}\right)-\frac{\kappa p_{0}}{L_{p} \rho_{0}} \frac{k_{\theta}^{2}}{\gamma_{\boldsymbol{k}}} \hat{\bar{\varphi}}_{\boldsymbol{k}}\left(k_{x}\right) \\
=&-v k_{x}^{4} \hat{\bar{\varphi}}_{\boldsymbol{k}}\left(k_{x}\right)-\frac{S}{\tau_{A}}\left|\tilde{b}_{r}\right|^{2} k_{x}^{2} \hat{\bar{\varphi}}_{\boldsymbol{k}}\left(k_{x}\right)-\frac{\kappa p_{0} \chi k_{\theta}^{2}}{\rho_{0} L_{p} \gamma_{\boldsymbol{k}}^{2}}  k_{x}^{2} \hat{\bar{\varphi}}_{\boldsymbol{k}}\left(k_{x}\right) \\
&-\left(\frac{S}{\tau_{A}}\right)^{2} \frac{R k_{\theta}^{2}}{L_{s}^{3}} \bar{\varphi}_{\boldsymbol{k}}(0) i \sqrt{2 \pi} \delta^{(1)}\left(k_{x}\right) I \\
&-\left(\frac{S}{\tau_{A}}\right)^{2} \frac{R k_{\theta}^{2}}{L_{s}^{3}} \bar{\varphi}_{\boldsymbol{k}}(0) r_{m n} \sqrt{2 \pi} \delta\left(k_{x}\right) I.
\end{aligned}
\label{finalequation}
\end{equation}
Since the stochastic magnetic field background is weak and the resultant turbulent viscosity $\nu$ is also small, we can treat the R.H.S. of Eq.(\ref{finalequation}) as a small perturbation. So Eq.(\ref{finalequation}) can be written as
\begin{equation}
\hat{H}_{0}\hat{\bar{\varphi}}_{\boldsymbol{k}}(k_{x})=\hat{H}_{1}\hat{\bar{\varphi}}_{\boldsymbol{k}}(k_{x}),
\label{concepteigen}
\end{equation}
where
\begin{displaymath}
\begin{aligned}
\hat{H}_{0}=-&\frac{S}{\tau_{A}} \frac{k_{\theta}^{2}}{L_{s}^{2}} \frac{d^{2}}{d k_{x}^{2}} +\gamma_{k} k_{x}^{2} -\frac{\kappa p_{0}}{L_{p} \rho_{0}} \frac{k_{\theta}^{2}}{\gamma_{k}} ,\\
\hat{H}_{1}=-&\nu k_{x}^{4} -\frac{S}{\tau_{A}}\left|\tilde{b}_{r}\right|^{2} k_{x}^{2} -\frac{\kappa p_{0} \chi k_{\theta}^{2}}{\rho_{0} L_{p} \gamma_{k}^{2}}  k_{x}^{2} \\
-&\left(\frac{S}{\tau_{A}}\right)^{2} \frac{R k_{\theta}^{2}}{L_{s}^{3}} i \sqrt{2 \pi} \delta^{(1)}\left(k_{x}\right) I\int d x' \delta(x')\mathcal{F}^{-1}\\
-&\left(\frac{S}{\tau_{A}}\right)^{2} \frac{R k_{\theta}^{2}}{L_{s}^{3}} r_{m n} \sqrt{2 \pi} \delta\left(k_{x}\right) I\int d x' \delta(x')\mathcal{F}^{-1}.
\end{aligned}
\end{displaymath}

We can then attain the corrected growth rate $\gamma_{\boldsymbol{k}}$ by doing perturbation theory. As equation $\hat{H}_{0}\hat{\bar{\varphi}}_{\boldsymbol{k}}^{(0)}(k_{x})=0$ is just the Fourier transform of Eq.(\ref{eigen}), the zeroth-order growth rate is just the growth rate given by Eq.(\ref{slow}). Thus the eigenmode solution of the ground state is
\begin{equation}
\hat{\bar{\varphi}}^{(0)}_{\boldsymbol{k}}\left(k_{x}\right)=\bar{\varphi}^{(0)}_{\boldsymbol{k}}(x=0)w_{\boldsymbol{k}} e^{-\frac{w_{\boldsymbol{k}}^{2} k_{x}^{2}}{2}},
\label{eigensolution}
\end{equation}
where $w_{\boldsymbol{k}}=\left(\frac{\tau_{A} \gamma_{k}^{(0)} L_{S}^{2}}{S k_{\theta}^{2}}\right)^{\frac{1}{4}}$.

By using perturbation theory, the first-order growth rate correction $\gamma_{\boldsymbol{k}}^{(1)}$ is given by the following equation:
\begin{equation}
\gamma_{\boldsymbol{k}}^{(1)}=\frac{\int_{-\infty}^{\infty} \hat{\bar{\varphi}}_{\boldsymbol{k}}^{(0)}\left(k_{x}\right) \hat{H}_{1} \hat{\bar{\varphi}}_{\boldsymbol{k}}^{(0)}\left(k_{x}\right) d k_{x}}{\int_{-\infty}^{\infty} \hat{\bar{\varphi}}_{\boldsymbol{k}}^{(0)}\left(k_{x}\right)\left[\partial_{\gamma_{\boldsymbol{{k}}}^{(0)}}\hat{H}_{0}\right] \hat{\bar{\varphi}}_{\boldsymbol{k}}^{(0)}\left(k_{x}\right) d k_{x}}.
\label{firstorderexpre}
\end{equation}
Plugging the expressions for $\hat{\bar{\varphi}}_{\boldsymbol{k}}^{(0)}$, $\hat{H}_{0}$ and $\hat{H}_{1}$ into Eq.(\ref{firstorderexpre}), the first-order correction to the growth rate of the ground state is
\begin{equation}
\gamma_{\boldsymbol{k}}^{(1)}=-\frac{5}{6}\hat{\nu}\left(\frac{\tau_{p}\tau_{\kappa}}{\tau_{A}^{2}}\right)^{\frac{1}{3}}S^{\frac{2}{3}}\tilde{k}_{\theta}^{\frac{2}{3}}-\frac{1}{3}\frac{S}{\tau_{A}}|\tilde{b}_{r}|^{2}
-\frac{2\sqrt{2}}{3}\frac{\hat{I}S^{\frac{4}{3}}\tilde{k}_{\theta}^{\frac{4}{3}}}{\left(\tau_{p}\tau_{\kappa}\tau_{A}^{4}\right)^{\frac{1}{3}}},
\label{firstorder}
\end{equation}
where
\begin{displaymath}
\hat{\nu}=\nu/L_{s}^{2}\quad \hat{I}=IR r_{mn}/L_s^{3}.
\end{displaymath}
Evidently, the first two terms of the expression for $\gamma_{\boldsymbol{k}}^{(1)}$ are negative definite, while the sign of the third depends on the sign of $I$. To determine whether $I$ is positive or negative, the turbulent viscosity $\nu$ should be calculated.

\subsubsection{\label{turbulentvis}Scaling of Turbulent Viscosity $\nu$}
Because $\nu$ originates from the $\boldsymbol{E}\times\boldsymbol{B}$ velocity fluctuation $\boldsymbol{\tilde{v}}$, it can be calculated through a simple nonlinear closure theory~\cite{Dupree1966,Garcia&Pat,Patbook}:
\begin{equation}
\nu=\sum_{\boldsymbol{k_{2}}}\left|\tilde{v}_{\boldsymbol{k_{2}}}\right|^{2} \tau_{\boldsymbol{k_{2}}},
\label{nuexpress}
\end{equation}
where $\tau_{\boldsymbol{k_{2}}}$ is the correlation time. A reasonable estimate of $\tau_{\boldsymbol{k_{2}}}$ is the reciprocal of the fast interchange growth rate $1/\gamma_{\boldsymbol{k_{2}}}^{(0)}$. As $\left|\tilde{v}_{\boldsymbol{k_{2}}}\right|=k_{2 \theta}\left|\tilde{\varphi}_{\boldsymbol{{k_{2}}}}\right|/B_{0}$, substituting Eq.(\ref{response}) into Eq.(\ref{nuexpress}), we get
\begin{equation}
\nu=\frac{R r_{mn}}{\pi^{-\frac{1}{2}}L_{s}^{5}} \left(\frac{S}{\tau_{A}}\right)^{2}\frac{\tilde{k}_{\theta}^{2}\bar{\varphi}_{\boldsymbol{k}}^{2}(0)}{B_{0}^{2}}\int d k_{2\theta}\frac{\left|k_{2\theta}\right|^{3}c^{2}Z^{2}w_{\boldsymbol{k_{2}}}o_{\boldsymbol{k_{2}}}^{2}}{\left(\Lambda_{\boldsymbol{k_{2}}}^{0}-\Lambda_{\boldsymbol{k_{2}}}\right)^{2}\gamma_{\boldsymbol{k_{2}}}^{(0)}},
\label{nu1}
\end{equation}
where both $\Lambda_{\boldsymbol{k}_{2}}^{0}$ and $\Lambda_{\boldsymbol{k}_{2}}$ are functions of $\nu$. So, to extract the scaling of $\nu$, we need a different approach. 

Recall in Eq.(\ref{microdrive}) and Eq.(\ref{green}), $\left(\Lambda_{\boldsymbol{k}_{2}}^{0}-\Lambda_{\boldsymbol{k}_{2}}\right)^{2}$ is equal to
\begin{equation}
\left(\Lambda_{\boldsymbol{k}_{2}}^{0}-\Lambda_{\boldsymbol{k}_{2}}\right)^{2}=\Bigg[\underbrace{\sqrt{\frac{2 \nu S k_{2 \theta}^{4}}{\tau_{A} L_{s}^{2}}}}_{\Lambda_{\boldsymbol{k}_{2}}^{0}}-\underbrace{\left(\frac{\kappa p_{0}}{\chi \rho_{0} L_{p}}-\nu k_{2 \theta}^{4}\right)}_{\Lambda_{\boldsymbol{k}_{2}}}\Bigg]^{2}.
\label{difference}
\end{equation}
Because of the fast interchange approximation, $\Lambda_{\boldsymbol{k_{2}}}^{0}=2\nu k_{2\theta}^{2}/w_{\boldsymbol{k_{2}}}^{2}\ll \nu k_{2\theta}^{4}$. In addition, in the weak-mean-pressure-gradient limit, i.e., $\tau_{p}\gg 1/\tau_{\kappa}\nu^{2}k_{2\theta}^{4}$, $\nu k_{2\theta}^{4}$ becomes the dominant term in the bracket of Eq.(\ref{difference}). The scaling of $\nu$ then is
\begin{equation}
\nu=\left[\pi^{\frac{1}{2}} \frac{R r_{mn}}{B_{0}^{2}}\frac{\tilde{k}_{\theta}^{2}}{L_{s}^{5}}\left(\frac{S}{\tau_{A}}\right)^{2}\bar{\varphi}_{\boldsymbol{k}}^{2}(0)\int d k_{2\theta}\frac{c^{2}Z^{2}w_{\boldsymbol{k_{2}}}o_{\boldsymbol{k_{2}}}^{2}}{\left|k_{2\theta}\right|^{5}\gamma_{\boldsymbol{k_{2}}}^{(0)}}\right]^{\frac{1}{3}}.
\label{scalinglaw}
\end{equation}
This limit can be justified by the following argument. 

If we retain the growth rate of $\tilde{\varphi}_{\boldsymbol{k_{2}}}$ and utilize the fast interchange approximation, Eq.(\ref{microdrive}) is modified to
\begin{equation}
\frac{\partial\tilde{\varphi}}{\partial t}+\lambda\tilde{\varphi}=\hat{D}\left[\tilde{b}_{r}\bar{\varphi}\right],
\label{FDT}
\end{equation}
where $\lambda=\nu k_{2\theta}^{2}-\left(1/\tau_{p}\tau_{\kappa}\right)^{1/2}$, and $\hat{D}\left[\tilde{b}_{r}\bar{\varphi}\right]$ denotes the drive by $\tilde{b}_{r}\bar{\varphi}$ beats. 
The point here is that $\tilde{\varphi}$ is subject to two drives: a linear drive by curvature and pressure gradient, which corresponds to the second term in $\lambda$, and a drive by the noise---i.e., $\hat{D}\left[\tilde{b}_{r}\bar{\varphi}\right]$. As indicated by Eq.(\ref{nuexpress}), $\nu$ will increase with $\left|\tilde{\varphi}\right|$, so $\tilde{\varphi}$ can not grow indefinitely, and there is a point at which $\nu$ becomes large enough that both drives saturate. In other words, \emph{the growth of $\tilde{\varphi}$ is over-saturated}. Over-saturation requires $\lambda>0$---i.e., $\nu k_{2\theta}^{2}>\left(1/\tau_{p}\tau_{\kappa}\right)^{1/2}$, which is consistent with the limit we used for Eq.(\ref{scalinglaw}). This is similar to the case of Ref.~\cite{Rame&PatPPCF}. Then Eq.(\ref{FDT}) immediately looks like a generalization of the Langevin equation~\cite{van1992}, which further implies a fluctuation-dissipation balance~\cite{sethna2021}. Here we can see the dual identities of $\boldsymbol{\tilde{b}}$: on the one hand, it serves as part of the noise to excite small-scale cells; on the other hand, the turbulent viscosity $\nu$ resulting from it damps small-scale cells. Therefore, as mentioned in Sec.~\ref{tildevarphi}, $\tilde{\varphi}$ and $\tilde{p}$ can reach equilibrium and be adiabatically modulated by the beat of $\boldsymbol{\tilde{b}}\tilde{\varphi}$. 

In addition, in this limit $\tau_{p}\gg 1/\tau_{\kappa}\nu^{2}k_{2\theta}^{4}$, the integral $I$ is positive, which means the first-order correction to the growth rate given by Eq.(\ref{firstorder}) is negative definite. \emph{So we can conclude the net effect of a stochastic magnetic field on the large-scale mode is to reduce its growth, in proportion to the magnitude of stochastic magnetic field intensity.}

\section{\label{analysis}Analysis: Effects of Stochastic Magnetic Field}
In the calculation of perturbed growth rate, we defined two operators $\hat{H}_{0}$ and $\hat{H}_{1}$, so as to divide the terms in Eq.(\ref{finalequation}) into two groups. All the terms involving the stochastic magnetic field  are put in $\hat{H}_{1}$, so that $\hat{H}_{0}\bar{\varphi}_{\boldsymbol{k}}(k_{x})=0$ is just the Fourier transform of Eq.(\ref{eigen}). Therefore, $H_{0}$ gives the zeroth-order growth rate of $\bar{\varphi}_{\boldsymbol{k}}$, which is consistent with the classical linear theory of the resistive interchange mode. Since the stochastic magnetic field is weak, $\hat{H}_{1}$ is regarded as a perturbation. To analyze the effects of stochastic magnetic field clearly, we can number the different terms of the expression for $\hat{H}_{1}$ as
\begin{equation}
\begin{aligned}
&\hat{H}_{1}=-\underbrace{\nu k_{x}^{4}}_{\textrm{\textcircled{1}}} -\underbrace{\frac{S}{\tau_{A}}\left|\tilde{b}_{r}\right|^{2} k_{x}^{2}}_{\textrm{\textcircled{2}}} -\underbrace{\frac{\kappa p_{0} \chi k_{\theta}^{2}}{\rho_{0} L_{p} \gamma_{k}^{2}}  k_{x}^{2}}_{\textrm{\textcircled{3}}}\\
-&\underbrace{\left(\frac{S}{\tau_{A}}\right)^{2} \frac{R k_{\theta}^{2}}{L_{s}^{3}}  \sqrt{2 \pi} \left[\delta^{(0)}r_{mn}+i\delta^{(1)}\right] I\int d k'_{x} \delta^{(0)}\mathcal{F}^{-1}}_{\textrm{\textcircled{4}}}.
\end{aligned}
\end{equation}

Among these four terms, the physics of term \textcircled{2}, which comes from the correlation $(a)$ in Eq.(\ref{mainequation}a), is clearest. Term \textcircled{2} shares the same form with the second term of the expression for $\hat{H}_{0}$, i.e., they are both quadratic functions of $k_{x}$. If we neglect the other perturbations, Eq.(\ref{finalequation}) is rewritten as
\begin{equation}
-\frac{S}{\tau_{A}} \frac{k_{\theta}^{2}}{L_{s}^{2}} \frac{d^{2}}{d k_{x}^{2}} \bar{\varphi}_{\boldsymbol{k}}+\left[\gamma_{\boldsymbol{k}}+\frac{S}{\tau_{A}}\left|\tilde{b}_{r}\right|^{2}\right] k_{x}^{2} \bar{\varphi}-\frac{\kappa p_{0}}{L_{p} \rho_{0}} \frac{k_{\theta}^{2}}{\gamma_{\boldsymbol{k}}} \bar{\varphi}_{\boldsymbol{k}}=0.
\label{revisedeigen}
\end{equation}
When the magnitude of the stochastic magnetic field is large, the corrected growth rate of the ground state is
\begin{equation}
\gamma_{\boldsymbol{k}}=\frac{\tilde{k}_{\theta}}{S\left|\tilde{b}_{r}\right|}\frac{\tau_{A}}{\tau_{p}\tau_{\kappa}}.
\label{vorticitydamping}
\end{equation}
In Eq.(\ref{vorticitydamping}), $\gamma_{\boldsymbol{k}}\propto\left|\tilde{b}_{r}\right|^{-1}$, so the larger the magnitude of the stochastic magnetic field, the smaller the growth rate of the large-scale mode.

If there is no stochastic field, the growth of $\bar{\varphi}_{\boldsymbol{k}}$ is driven against inertia by torque produced by pressure gradient. Now, the stochastic magnetic field effectively adds to the inertia of plasma and thus stabilizes the growth of mode. This effect is magnetic vorticity damping. By re-expressing
\begin{equation}
\frac{S}{\tau_{A}}\left|\tilde{b}_{r_{\boldsymbol{k_{2}}}}\right|^{2}=\frac{v_{A}^{2}}{\eta}\frac{k_{2\theta}^{2}}{L_{s}^{2}}o_{\boldsymbol{k_{2}}}^{4},
\label{re-express}
\end{equation}
we can balance it with the linear bending term and obtain
\begin{equation}
\begin{aligned}
\frac{v_{A}^{2}}{\eta}\frac{k_{2\theta}^{2}}{L_{s}^{2}}\frac{o_{\boldsymbol{k_{2}}}^{4}}{(\Delta x)^{2}}&\bar{\varphi}_{\boldsymbol{k_{2}}}\sim\frac{v_{A}^{2}}{\eta}\frac{k_{\theta}^{2}}{L_{s}^{2}}(\Delta x)^2\bar{\varphi}_{\boldsymbol{k_{2}}}\\
o_{\boldsymbol{k_{2}}}\sim&\left[\frac{k_{\theta}^{2}}{k_{2\theta}^{2}}(\Delta x)^{4}\right]^{\frac{1}{4}}.
\end{aligned}
\label{balancing}
\end{equation}
Eq.(\ref{balancing}) offers us a criterion when magnetic vorticity damping becomes a significant effect, and reduces the growth of the mode. This result is a reminiscent of Rutherford's 1973 work on the tearing mode~\cite{Rutherford1973}. In that paper, the perturbed magnetic field growing with growth rate $\gamma$ can induce a perturbed current, which further produces a torque that can drive the tearing mode against plasma inertia. But as the perturbed field grows, the nonlinear force will gradually dominate and produce a torque opposing the growth of the mode. By balancing the torque produced by linear and nonlinear forces, Rutherford noted the system enters the nonlinear regime when the widths of the magnetic islands become comparable to the width of the tearing layer, i.e., when $o_{\boldsymbol{k_{2}}}\sim\Delta x$. In our model, this corresponds to the condition when Eq.(\ref{balancing}) holds. The stochastic magnetic field resembles the nonlinear force in Rutherford's model. The difference between Rutherford's model and ours is also significant: the extra factor $(k_{\theta}^{2}/k_{2\theta}^{2})$ in Eq.(\ref{balancing}) reflects the multi-scale nature of this problem.

Term \textcircled{1} and term \textcircled{3} are both related to turbulent viscosity $\nu$ (or turbulent diffusivity $\chi$, which is equal to $\nu$ in this model). As discussed in Sec.~\ref{introstoch} and Sec.~\ref{turbulentvis}, those small-scale convective cells produce a turbulent viscosity $\nu$. So it is not a surprise to find that the mode is stabilized by this turbulent viscosity (the first term on the R.H.S of Eq.(\ref{firstorder}) is negative).

The physics of term \textcircled{4} is more complex. It originates from the correlation $(b)$ in Eq.(10a)(BTW, correlation $(c)$ vanishes since it is a total derivative). We can rewrite correlation $(b)$ as
\begin{equation}
\left\langle\nabla_{\|}^{(0)} \tilde{\boldsymbol{b}} \cdot \nabla_{\perp} \tilde{\varphi}\right\rangle\sim\left\langle\nabla_{\|}^{(0)} \tilde{b}_{r} \tilde{E}_{r}\right\rangle\sim\nabla_{\|}^{0}\left\langle\tilde{J}_{\|^{0}}\right\rangle^{(3)}.
\label{radialcurrent}
\end{equation}
From Eq.(\ref{radialcurrent}) we can see, the potential fluctuation produces a fluctuating radial electric field, which further generates a current parallel to $\boldsymbol{B_{0}}$. In the final result, the correction to the growth rate from term \textcircled{4} is the last term in Eq.(\ref{firstorder}), whose sign depends on the sign of $I$. Since we have taken the limit $L_{p}\gg c_{s}/\tau_{\kappa}\nu^{2}k_{2\theta}^{4}$, $I$ is positive, which means term \textcircled{4} can also reduce the growth of the mode. This is due to the fact that large-scale mode is electrostatically scattered by small-scale convective cells.

In summary, the interaction between the large-scale mode and small-scale convective cells forms a feedback loop. As illustrated in the FIG.~\ref{flowchart}, the stochastic magnetic field and the large-scale cell together can drive small-scale cells while small-scale cells react on the large-scale cell through two different approaches: electrostatic scattering and turbulent viscosity.

\section{\label{conclusion}Conclusion and Discussion}
In this paper, we presented an in-depth analysis of the theory of instability and turbulent relaxation in a stochastic magnetic field. For tractability, we focus on a comparatively simple, yet relevant and representative, system---namely that of the electrostatic resistive interchange. Here, the static magnetic fluctuations which underpin the stochasticity render parallel gradients $\nabla_{\|}\rightarrow\nabla_{\|}^{(0)}+\boldsymbol{\tilde{b}}\cdot\nabla_{\bot}$, and so modify the basic structure of the eigenmode equation by converting it to a stochastic differential equation. This is, in turn, solved by the method of averaging, which exploits the scale separation between the low-$\boldsymbol{k}$ resistive interchange test mode, and the small-scale magnetic perturbations. The resulting dynamics are intrinsically multi-scale. Our study yields both general results---applicable to any instance of instability in a stochastic background---and quantitative results specific to this problem. 

The broadly applicable findings of this paper are:
\begin{enumerate}[i.)]
\item maintaining quasi-neutrality ($\nabla\cdot\boldsymbol{J}=0$) at all orders reveals that electrostatic convective cell turbulence is driven at small scales by the beat of small-scale magnetic perturbations $\boldsymbol{\tilde{b}}$ and large-scale mean electrostatic potential $\bar{\varphi}$---i.e., via $\boldsymbol{\tilde{b}}\bar{\varphi}$ modulation. This effectively converts the problem to one of turbulent dynamics, and tells as that turbulence with small-scale structure is generated.
\item the small-scale turbulence in turn modifies the large-scale mode via an effective flow viscosity and thermal diffusivity (computed by closure), as well as electrostatic scattering which is given by correlation $(b)$ in Eq.(\ref{mainequation}a). Thus, the dynamics take on the character of a disparate scale interaction, with large scale $\rightarrow$ small scale modulations and feedback by small scale $\rightarrow$ large scale scattering
\item the stochastic magnetic perturbations produce a magnetic braking effect, which exerts a drag on large-scale vorticity. This effect is similar in structure to the nonlinear $\boldsymbol{J}\times\boldsymbol{B}$ force identified by Rutherford, but in our case it is produced by the stochastic magnetic perturbations. 
\item the generation of small-scale cells due to $\boldsymbol{\tilde{b}}\bar{\varphi}$ interaction implies that correlation develops between the electrostatic turbulence and the ambient stochastic field---i.e. $\langle\tilde{b}_{r}\tilde{v}_{r}\rangle\neq0$ is shown. Here $\boldsymbol{\tilde{v}}$ refers to the small-scale cell velocity. Thus, we see that the velocity fluctuations `lock on' to the ambient static magnetic perturbations. This will necessarily affect the statistics of the turbulence.
\end{enumerate}
We anticipate that results i.)--iv.) will be of broad interest in the context of RMP experiments. 

The specific detailed calculations of this paper are:
\begin{enumerate}[i.)]
\item the net effect of stochastic magnetic fields is to reduce resistive interchange growth---i.e., a trend toward stabilization. The increment is calculated in Eq.(\ref{firstorder}). Note this result is contrary to previous ones, and is a consequence of vorticity damping and diffusion emerging as the principal effects.
\item the turbulent viscosity and turbulent thermal diffusivity driven by the small-scale convective cells are calculated. The specific result is given by Eq.(\ref{scalinglaw}), with $\nu=\chi$.
\item the width of the magnetic islands when the magnetic braking effect becomes significant is calculated and given by Eq.(\ref{balancing}). This differs from Rutherford's result by a factor of $k_{\theta}^{2}/k_{2\theta}^{2}$, on account of the multi-scale nature of the problem considered here.
\item the $\langle\tilde{b}_{r}\tilde{v}_{r}\rangle$ correlation is calculated explicitly, and given by Eq.(\ref{bvcorrelation}).
\end{enumerate}
Taken together, these results, \emph{which constitute computationally testable predictions}, fully characterize the state of the system. 

\begin{figure}[h]
    \centering
    \includegraphics[scale=1]{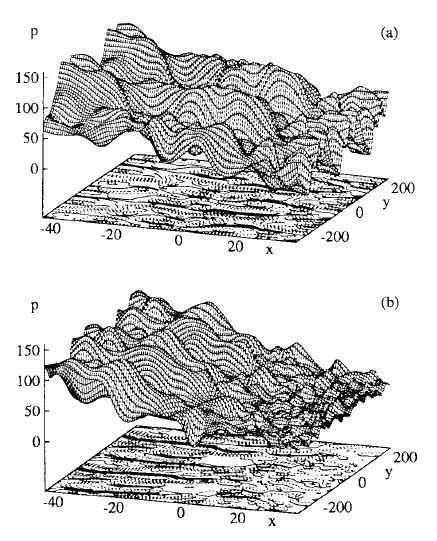}
    \caption{Plasma pressure in a sector at the low field side without (a) and with RMP (b). Clearly large-scale structures are suppressed in the stochastic layer, and spatial roughness increases.}
    \label{Beyer}
\end{figure}

Previous simulation work has addressed this subject. Of particular note is the paper by Beyer, et al., which described a study of electrostatic resistive ballooning modes in a background stochastic magnetic field~\cite{Beyer1998}. FIG.~\ref{Beyer} is a result of that study, 
\noindent and contrasts the pressure fluctuation profile in a smooth field with one in a stochastic field. The latter clearly manifests increased small-scale structure and spatial roughness. These are consistent with our findings that small-scale convective cells will be generated by the interaction of $\boldsymbol{\tilde{b}}$ with large-scale mode. The Beyer, et al. study did not analyze this aspect of the results in detail. We have suggested that an interesting continuation and a rather precise test of our theory would be a measurement of the correlation between $\boldsymbol{\tilde{b}}$ and small-scale $\tilde{v}_{r}$---i.e., $\langle\tilde{b}_{r}\tilde{v}_{r}\rangle$ and a comparison to the prediction in Eq.(\ref{bvcorrelation}). Similarly, a comparison of the turbulent flux $\left\langle\tilde{v}_{r}\tilde{p}\right\rangle$ and diffusivity $\chi$ driven at small scales with the prediction of Eq.(\ref{scalinglaw}) would be of considerable interest.

Experimental studies of such fine scale fluctuation dynamics are understandably challenging. An interesting and relevant result was recently reported by Choi et al.~\cite{Choi2021}, who compared the change in pedestal temperature fluctuation predictability (as deduced from Jensen-Shannon entropy) with RMP switched on and off. These studies focused on the putative stochastic region at the edge of RMP-induced islands in the pedestal. Results indicate that the effect of stochasticity is to reduce the Jensen-Shannon complexity~\cite{Rosso2007} and predictability of the pedestal turbulence. One possible cause of this change would be that correlations between the turbulence and stochastic field develop---i.e., $\langle\tilde{b}_{r}\tilde{v}_{r}\rangle\neq 0$---as predicted here. The related generation of small-scale structure, as we suggest, is another possible cause of the drop in predictability. Of course, this finding is not entirely surprising, as it is well known that external noise can suppress or inhibit the instability characteristic of chaotic orbits~\cite{Hirsh1982}. Interestingly, however, Choi et al. also report an increase in bicoherence in the pedestal turbulence. We suggest that this may be due to the generation of small-scale cells, which can increase spectral transfer. A possible next step is to determine the change in the measure of fluctuation complexity implied by our results, and to compare this with the experimental findings.

In addition to the suggestions listed above, several other avenues for future research have been identified. One---of particular relevance to tokamak applications---is to consider twisted slicing mode~\cite{twisted1965} (or equivalently ballooning mode~\cite{CHT1978}) structure in a stochastic magnetic field. These modes may be thought of as extending along magnetic field lines, which now wander, stochastically. This points towards a natural critical competition between the field line decorrelation length (i.e., the counterpart of the Lyapunov exponent~\cite{Rech&Rosen&White1979}) and the extent of the mode along the field line. A second topic is, of course, a turbulent large-scale state, as opposed to a singe mode case. Here, the presence of the stochastic field and the modulationally generated small-scale convective cells potentially open the possibility of increased nonlinear transfer, by increasing the number of triad interactions. This offers the possibility of reconciling the decrease in complexity/predictability observed in Ref.~\cite{Choi2021} with the increase in bicoherence also observed. Finally, extension of this analysis to a kinetic description of microinstabilities should be considered. Here, since $v_{\|}\nabla_{\|}\rightarrow v_{\|}(\nabla_{\|}^{(0)}+\boldsymbol{\tilde{b}}\cdot\nabla_{\bot}))$, ambient stochastic perturbations will scatter particle streaming. These topics will be studied in future publications. 

\begin{acknowledgments}
We thank Minjun Choi, Xavier Garbet, Min Xu, Jae-Min Kwon, Zhibin Guo and T.S. Hahm for stimulating conversations. Mingyun Cao thanks Shanghai Jiao Tong University for his undergraduate education, where this work began as part of his Bachelor Thesis, and also thanks Peking University for hospitality while some of this work was performed. Patrick Diamond acknowledges participants in the 2019 Festival de Théorie and the 2021 Festival Lectures for discussions. This research was supported by the U.S. Department of Energy under Award Number DE-FG02-04ER54738.
\end{acknowledgments}
\appendix

\section{\label{s&f}Calculations of the growth rate of ``ground state" under slow and fast interchange ordering approximations}
The normalized eigen solutions to Eq.(\ref{eigen}) are
\begin{equation}
\varphi_{\boldsymbol{k}}^{j}=\frac{\alpha_{\boldsymbol{k}}^{1/2}}{\pi^{1/4}}\frac{1}{\sqrt{2^{j}j!}}H_{j}\left(\alpha_{\boldsymbol{k}}x\right)e^{-\frac{\left(\alpha_{\boldsymbol{k}}x\right)^{2}}{2}},
\label{exactsol}
\end{equation}
where
\begin{displaymath}
\alpha_{\boldsymbol{k}}=\left(\frac{Sk_{\theta}^{2}}{\gamma_{\boldsymbol{k}}\tau_{A}L_{s}^{2}}\right)^{\frac{1}{4}},
\end{displaymath}
and its corresponding growth rate $\gamma_{\boldsymbol{k}}$ satisfies
\begin{displaymath}
\left(2j+1\right)\sqrt{\frac{\gamma_{\boldsymbol{k}}Sk_{\theta}^{2}}{\tau_{A}L_{s}}}=\left(-\gamma_{\boldsymbol{k}} k_{\theta}^{2}+\frac{\kappa p_{0} k_{\theta}^{2}}{\rho_{0} L_{p} \gamma_{\boldsymbol{k}}}\right).
\end{displaymath}
This is equation is not easy to solve, but in the following two limits, we can get $\gamma_{\boldsymbol{k}}$ easily.
\subsection{slow interchange ordering}
For slow interchange, $k_{r}$ is much larger than $k_{\theta}$, so the term $\gamma_{\boldsymbol{k}}k_{\theta}^{2}\varphi_{\boldsymbol{k}}$ in Eq.(\ref{eigen}) can be neglected. Then Eq.(\ref{eigen}) reduces to
\begin{equation}
-\gamma_{\boldsymbol{k}} \frac{\partial^{2} \varphi_{\boldsymbol{k}}}{\partial x^{2}}+\frac{S}{\tau_{A}} \frac{k_{\theta}^{2}}{L_{S}^{2}} x^{2} \varphi_{\boldsymbol{k}}-\frac{\kappa p_{0} k_{\theta}^{2}}{\rho_{0} L_{p} \gamma_{\boldsymbol{k}}} \varphi_{\boldsymbol{k}}=0.
\end{equation}
In this condition, the growth rate of the "ground state" is
\begin{displaymath}
\gamma_{\boldsymbol{k}}=S^{-\frac{1}{3}}\tau_{A}^{\frac{1}{3}} \tau_{p}^{-\frac{2}{3}} \tau_{\kappa}^{-\frac{2}{3}} \tilde{k}_{\theta}^{\frac{2}{3}},
\end{displaymath}
which is exactly the Eq.(\ref{slow}).

\subsection{fast interchange ordering}
For fast interchange, $k_{\theta}$ is much larger than $k_{r}$, so the bending term and first term of Eq.(\ref{eigen}) can be neglected, which means we just need to balance the last two terms. Then we obtain
\begin{displaymath}
\gamma_{\boldsymbol{k}}=\tau_{p}^{-\frac{1}{2}} \tau_{\kappa}^{-\frac{1}{2}},
\end{displaymath}
which is exactly the Eq.(\ref{fast}). Here we notice that growth rate of fast interchange is independent of $k_{\theta}$.

\section{\label{KP}Kadomtsev and Pogutse's Model}
In K\&P's work, they calculated the radial electron heat flux in a stochastic magnetic field. Originally, the heat flux is
\begin{equation}
\boldsymbol{q}=-\chi_{\parallel}\nabla_{\|}T-\chi_{\bot}\nabla_{\bot}T,
\end{equation}
where $\chi_{\|}$ and $\chi_{\bot}$ are longitudinal and transverse thermal conductivity respectively, and $\chi_{\|}\gg\chi_{\bot}$. Now with stochastic magnetic field, the heat flux becomes
\begin{equation}
\boldsymbol{q}=-\chi_{\|}\left(\nabla_{\|}^{(0)}+\tilde{\boldsymbol{b}} \cdot \nabla\right)(\bar{T}+\tilde{T})\left(\boldsymbol{b}_{\mathbf{0}}+\tilde{\boldsymbol{b}}\right)-\chi_{\perp} \nabla_{\perp}(\bar{T}+\tilde{T}),
\end{equation}
and the heat flux fluctuation $\tilde{\boldsymbol{q}}$ is
\begin{equation}
\tilde{\boldsymbol{q}}=-\chi_{\perp} \nabla_{\perp} \tilde{T}-\chi_{\|}\left(\nabla_{\|}^{(0)} \tilde{T}+\tilde{\boldsymbol{b}} \cdot \nabla \bar{T}\right) \boldsymbol{b}_{\mathbf{0}}.
\end{equation}
N.B. $\bar{T}$ is only a function of $x$.

Since $\nabla\cdot\boldsymbol{q}=0$ at all scales, $\tilde{\boldsymbol{q}}$ should also be divergence-free, which gives us
\begin{equation}
-\chi_{\|} \nabla_{\|}^{(0)} \tilde{T}-\chi_{\perp} \nabla_{\perp}^{2} \tilde{T}=\chi_{\|} \nabla_{\|}^{(0)}(\tilde{\boldsymbol{b}} \cdot \nabla \bar{T}).
\end{equation}
Therefore, the response of $\tilde{T}$ to $\tilde{\boldsymbol{b}}$ is
\begin{equation}
\tilde{T}_{\boldsymbol{k}}=-\frac{\chi_{\|} i k_{\|} \tilde{b}_{{r}_{\boldsymbol{k}}} }{\chi_{\|} k_{\|}^{2}+\chi_{\perp} k_{\perp}^{2}}\frac{\partial(T)}{\partial r},
\end{equation}
and the mean nonlinear radial flux is
\begin{equation}
\begin{aligned}
\left\langle q_{r}\right\rangle_{N L}&=-\chi_{\|}\langle\tilde{b}_{r}\widetilde{\boldsymbol{b}\cdot\nabla T}\rangle\\
&=-\chi_{\|} \frac{\partial\langle T\rangle}{\partial r} \sum_{\boldsymbol{k}} \frac{\chi_{\perp} k_{\perp}^{2}\left|b_{r_{\boldsymbol{k}}}\right|^{2}}{\chi_{\|} k_{\|}^{2}+\chi_{\perp} k_{\perp}^{2}}\\
&=-\sqrt{\chi_{\|} \chi_{\perp}}\left\langle\tilde{b}^{2}\right\rangle l_{a c}\left\langle\sqrt{k_{\perp}^{2}}\right\rangle \frac{\partial\langle T\rangle}{\partial r},
\end{aligned}
\end{equation}
where $l_{ac}$ is the auto-correlation length of the stochastic magnetic field.
Immediately, we can see there is a comparison relation between K\&P's and C\&D's model, as is listed in table.(\ref{table1}).
\begin{table}[H]
\centering
\caption{\label{table1}
Comparison between K\&P's and C\&D's Models
}
\begin{ruledtabular}
\begin{tabular}{cccc}
Analogy&
K\&P&
C\&D\\
\colrule
Goal & $\langle q_{r}\rangle_{NL}$ & $\gamma_{\boldsymbol{k}}^{(1)}$\\
Base State & $\bar{T}$ & $\bar{\varphi}$\\
Stochastic quantity & $\tilde{\boldsymbol{b}}$ & $\tilde{\boldsymbol{b}}$\\
Constraint & $\nabla\cdot\boldsymbol{q}=0$ & $\nabla\cdot\boldsymbol{J}=0$\\
Resulting Fluctuations & $\tilde{T}$ & $\tilde{\varphi}$\\
\end{tabular}
\end{ruledtabular}
\end{table}

% The \nocite command causes all entries in a bibliography to be printed out
% whether or not they are actually referenced in the text. This is appropriate
% for the sample file to show the different styles of references, but authors
% most likely will not want to use it.
%\nocite{*}
\bibliography{aipsamp}% Produces the bibliography via BibTeX.

\end{document}